\documentclass[manuscript]{aastex}

\usepackage{txfonts}
\usepackage{longtable}
\usepackage{rotating}
\usepackage{natbib}
\usepackage{graphicx}
\usepackage{graphics}
\usepackage{psfrag}
\usepackage{amssymb}
\bibpunct{(}{)}{;}{a}{}{,}
\def\Teff{$T_{\mathrm{eff}}$}
\def\logg{\ensuremath{\log g}}
\def\vmic{$\upsilon_{\mathrm{mic}}$}
\def\vmac{$\upsilon_{\mathrm{macro}}$}
\def\vsini{\ensuremath{{\upsilon}\sin i}}
\def\kms{$\mathrm{km\,s}^{-1}$}

\def\exc{$\chi_{\mathrm{excit}}$}
\def\loggf{log $gf$}

\def\espa{ESPaDOnS}

\def\nlte{non-LTE}
\def\llm{{\sc LLmodels}}
\def\hfs{{\it hfs}}

\def\synth{{\sc SYNTH3}}
\def\vald{{\sc VALD}}

\def\logl{\ensuremath{\log L/L_{\odot}}}

\def\M{\ensuremath{M/M_{\odot}}}

\shorttitle{A spectropolarimetric analysis of WASP-12}
\shortauthors{Fossati et al.}

\begin{document}

\title{A detailed spectropolarimetric analysis of the planet hosting star
WASP-12}
\altaffiltext{1}{Based on observations obtained at the Canada-France-Hawaii
Telescope (CFHT), which is operated by the National Research Council of
Canada, the Institut National des Sciences de l'Univers of the Centre
National de la Rechereche Scientifique of France, and the University of
Hawaii.}
\altaffiltext{2}{Based on observations made with the NASA/ESA Hubble Space
Telescope, obtained MAST at the Space Telescope Science Institute, which is
operated by the Association of Universities for Research in Astronomy, Inc.,
under NASA contract NAS 5-26555. These observations are associated with
program \#11651.}

\author{L. Fossati}
\affil{Department of Physics and Astronomy, Open University,
	Walton Hall, Milton Keynes MK7 6AA, UK}
\email{l.fossati@open.ac.uk}
\and
\author{S. Bagnulo}
\affil{Armagh Observatory, College Hill, Armagh BT61 9DG, Northern Ireland, UK}
\email{sba@arm.ac.uk}
\and
\author{A. Elmasli\altaffilmark{3} and C.~A. Haswell and S. Holmes}
\affil{Department of Physics and Astronomy, Open University,
	Walton Hall, Milton Keynes MK7 6AA, UK}
\email{elmasli@ankara.edu.tr,C.A.Haswell@open.ac.uk,s.holmes@open.ac.uk}
\and
\author{O. Kochukhov}
\affil{Department of Physics and Astronomy, Uppsala University, SE-751 20,
	Uppsala, Sweden}
\email{Oleg.Kochukhov@fysast.uu.se}
\and
\author{E.~L. Shkolnik}
\affil{Department of Terrestrial Magnetism, Carnegie Institution of
	Washington, 5241 Broad Branch Road, NW, Washington, DC 20015, USA}
\email{shkolnik@dtm.ciw.edu}
\and
\author{D.~V. Shulyak}
\affil{Institute of Astrophysics, Georg-August-University,
	Friedrich-Hund-Platz 1, D-37077, G\"ottingen, Germany}
\email{denis.shulyak@gmail.com}
\and
\author{D. Bohlender}
\affil{Herzberg Institute of Astrophysics, National Research Council of
	Canada, 5071 West Saanich Road, Victoria, BC V9E 2E7, Canada}
\email{david.bohlender@nrc-cnrc.gc.ca}
\and
\author{B. Albayrak}
\affil{Department of Astronomy and Space Sciences, Ankara
	University, 06100, Tando\u{g}an, Ankara, Turkey}
\email{balbayrak@ankara.edu.tr}
\and
\author{C. Froning}
\affil{Center for Astrophysics and Space Astronomy, University of Colorado,
	593 UCB, Boulder, CO 80309-0593, USA}
\email{cynthia.froning@colorado.edu}
\and
\author{L. Hebb}
\affil{Department of Physics and Astronomy, Vanderbilt University, 6301
	Stevenson Center Nashville, TN 37235, USA}
\email{leslie.hebb@vanderbilt.edu}

\altaffiltext{3}{Department of Astronomy and Space Sciences, Ankara
	University, 06100, Tando\u{g}an, Ankara, Turkey}

\begin{abstract}
The knowledge of accurate stellar parameters is paramount in several fields of
stellar astrophysics, particularly in the study of extrasolar planets, where
often the star is the only visible component and therefore used to
infer the planet's fundamental parameters. Another important aspect of the
analysis of planetary systems is the stellar activity and the possible
star-planet interaction. Here we present a
self-consistent abundance analysis of the planet hosting star WASP-12 and a
high-precision search for a structured stellar magnetic field on the basis
of spectropolarimetric observations obtained with the \espa\ 
spectropolarimeter. Our results show that the star does not have a structured
magnetic field, and that the obtained fundamental parameters are in good
agreement with what was previously published. In addition we derive improved
constraints on the stellar age (1.0--2.65\,Gyr), mass (1.23--1.49\,\M),
and distance (295--465\,pc). WASP-12 is an ideal object to look for pollution
signatures in the stellar atmosphere. We analyse the WASP-12 abundances as a
function of the condensation temperature and compare them with those
published by several other authors on planet hosting and non-planet hosting
stars. We find hints of atmospheric pollution in WASP-12's photosphere,
but are unable to reach firm conclusions with our present data. We conclude
that a differential analysis based on WASP-12 twins will probably
clarify if an atmospheric pollution is present, the nature of this pollution
and its implications in the planet formation and evolution. We
attempt also the direct detection of the circumstellar disk through infrared
excess, but without success.
\end{abstract}


\section{Introduction}\label{intro}
One of the biggest surprises in the exoplanet field was the discovery of gas
giant planets orbiting very close to their host star. These hot Jupiter
planets represent one extreme of the Galaxy's population of planets, and
they provide important constraints to guide our nascent ideas about the
formation and evolution of planetary systems. The probability of transit
for close-in giant planets is $\sim 10\%$ \citep{seager2000}, and through
the analysis of transit light curves the ratio of the stellar and planetary
radii can be deduced. Through the significant uncertainties in the mass and
radius of any particular star, our characterisation of exoplanets is limited
by that of their host stars \citep[e.g.][]{southworth}. For this reason, it is
important to directly measure the properties of planet hosting stars,
particularly in cases where we expect the presence of planets may have
influenced the properties of the star through star-planet interactions.

One of the most extreme hot Jupiter exoplanets is WASP-12\,b, a gas giant
planet orbiting only 0.023\,AU from a late F-type host star \citep{hebb}.
WASP-12\,b's orbit is, therefore, only about 1.5 stellar diameters from the
photosphere of the star. At such proximity, interactions between the star
and the planet must occur. Near-UV observations of WASP-12 covering the
wavelengths of many resonance lines reveal that WASP-12\,b is surrounded by
an exosphere which appears to overfill its Roche lobe \citep{fossati2010}.
This exospheric gas may be the consequence of tidal disruption of the
planet's convective envelope as recently suggested by \citet{li2010}, but
it could also be entrained material from the stellar corona.

A planet orbiting as close as WASP-12\,b might be expected to interact
magnetically with its host star, c.f. \citet{es03}, \citet{es05}, and
\citet{es08}. A first step to search for such interactions is to detect
and quantify the stellar magnetic field by spectropolarimetry, e.g.
\citet{fares}. The presence of a magnetic field belonging to WASP-12 would
provide a precious piece of information needed to establish which mechanism
controls the structure and the evolution of the disk \citep{ch2010}.

The solar system's giant planets have enhanced metal abundances relative
to the Sun \citep{guillot05} and high atmospheric metal abundances have
been suggested as a contributing factor in the inflated radii of planets
such as WASP-12\,b \citep{burrows07}. Intriguingly, \citet{fossati2010}
detected a wealth of metallic atoms and ions in the exospheric gas
surrounding WASP-12\,b. If this gas is indeed accreting onto the host star
as suggested by \citet{li2010}, this could lead to abundance anomalies in the
photosphere of WASP-12. Since WASP-12 is expected to have a very shallow
surface convection zone, any accreted gas will remain close to the surface
rendering any pollution of the surface composition relatively easy to detect.

In this paper, we report on spectropolarimetric observations of WASP-12 which
we use to probe the stellar magnetic field, fundamental parameters and
abundance pattern of the star. In Sect.~\ref{obs} we describe our observations
and data reduction. Sect.~\ref{model} and Sect.~\ref{parameters} provide a
description of the adopted model atmosphere including methods and results of
the stellar parameter determination and abundance analysis. In
Sect.~\ref{mag.field} we provide the results of our stellar magnetic field
search. Our results are finally discussed in Sect.~\ref{discussion}, while in
Sect.~\ref{conclusions} we gather our conclusions.
\section{Observations and data reduction}\label{obs}
We observed WASP-12 using the \espa\  (Echelle SpectroPolarimetric Device
for ObservatioNs of Stars) spectropolarimeter at the Canada-France-Hawaii
Telescope (CFHT) on the 3rd and 5th of January 2010. The observations
were performed in ``polarimetric" mode.

\espa\ consists of a table-top cross-dispersed echelle spectrograph fed via
a double optical fiber directly from a Cassegrain-mounted polarisation
analysis module. Beside the natural intensity $I$, in polarimetric mode the
instrument can acquire a Stokes $V$ (or $Q$ or $U$) profile throughout the
spectral range 3700--10400\,\AA\ with a resolving power of about 65\,000.
A complete polarimetric observation consists of a sequence of 4 sub-exposures
\citep{donatietal1997,wade2000}.

Each of the four sub-exposures was 1290\,seconds long, with a total amount
of integration time of 1.5\,hrs, each night. The spectra were reduced
using the Libre-ESpRIT
package\footnote{\tt www.ast.obs-mip.fr/projets/espadons/espadons.html}
\citep{donati2007}. The Stokes $I$ spectra have a signal-to-noise ratio
(SNR) per pixel of $\sim$126 and $\sim$158 in the continuum, on the first and
second nights respectively; both values calculated at 5000\,\AA. To increase
the SNR of the Stokes $I$ spectrum we averaged the two available spectra,
obtaining a single spectrum with a SNR of $\sim$200, normalised by fitting
a low order polynomial to carefully selected continuum points.

The effective temperature (\Teff) was determined (see Sect.~\ref{parameters})
from our \espa\ data and two spectra of WASP-12 obtained at the
Isaac Newton Telescope (INT) with the Intermediate Dispersion Spectrograph
(IDS). The spectra cover the region of the H$\alpha$ line with a spectral
resolution of $R$=8\,000 \citep[see][for more details]{hebb}.

We also used Near-UV observations obtained with the HST Cosmic Origin
Spectrograph (COS) \citep{green2010,osterman2010} for the analysis
of the spectral energy distribution. The spectra, calibrated in flux, cover
three non-contiguous wavelength ranges in the Near-UV with a resolution of
$R\sim$20\,000. These observations are described in detail in
\citet{fossati2010}.
\section{The model atmosphere}\label{model}
To compute model atmospheres of WASP-12 we employed the \llm\ stellar model
atmosphere code \citep{llm}. For all the calculations Local Thermodynamical
Equilibrium (LTE) and plane-parallel geometry were assumed. We used the
\vald\ database \citep{vald1,vald2,vald3} as a source of atomic line
parameters for opacity calculations. The recent VALD compilation contains
information for about $6.6\times10^7$ atomic transitions, most of them coming
from the latest theoretical calculations performed by
R.~Kurucz\footnote{{\tt http://kurucz.harvard.edu}}. Convection was
implemented according to the \citet{cm1,cm2} model of convection
\citep[see][for more details]{heiter}.
\section{Fundamental parameters and abundance analysis}\label{parameters}
\citet{hebb} derived the fundamental parameters of WASP-12 from the analysis 
of low and mid-resolution spectra, obtaining \Teff=6300$\pm$150\,K,
\logg=4.38$\pm$0.10, and adopting a value of 0.85\,\kms\ for the
microturbulence velocity (\vmic). We used these values as our starting point
in an iterative process to gradually improve the parameters using different
spectroscopic indicators. In our analysis, every time any of \Teff, \logg,
\vmic\ or abundances changed during the iteration process, we recalculated
a new model with the implementation of the last measured quantities.
Similarly the derived abundances were treated iteratively: while the results
of the abundance analysis depend upon the assumed model atmosphere, the
atmospheric temperature-pressure structure itself depends upon the adopted
abundances, so we recalculated the model atmosphere every time abundances
were changed, even if the other model parameters were fixed. This ensures
the model structure is consistent with the assumed abundances.

\begin{figure}
\begin{center}
\includegraphics[width=\hsize,clip]{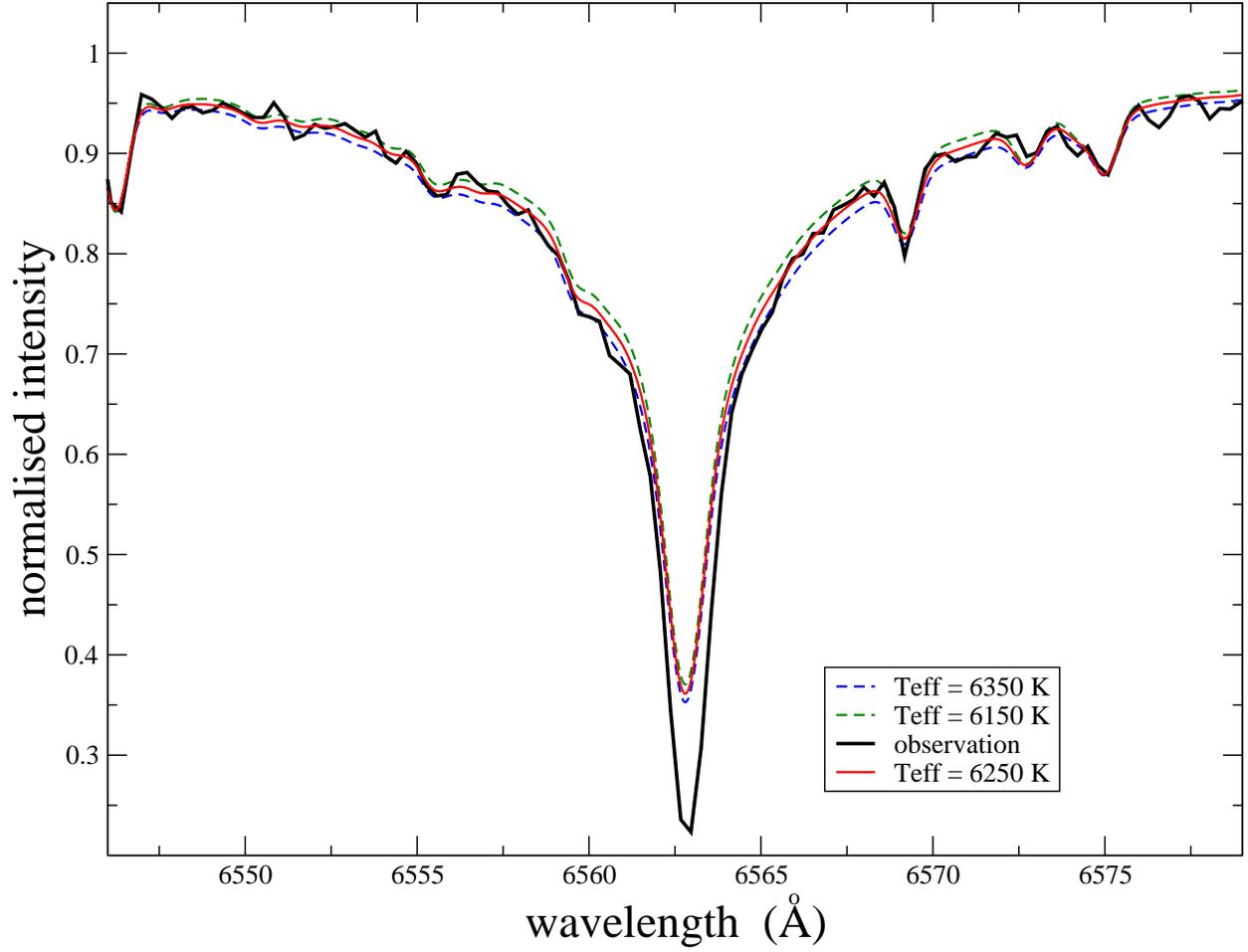}
\caption{Comparison between the H$\alpha$ line profile observed with the IDS
spectrograph (black solid line) and synthetic profiles calculated with the
final adopted \Teff\ of 6250\,K (red solid line), and uncertainty
(dashed lines).}
\label{hydrogen}
\end{center}
\end{figure}

We determined \Teff\ by fitting synthetic line profiles,
calculated with \synth\ \citep{synth3}, to the observed profiles of two
hydrogen lines: H$\alpha$ (from the IDS spectrograph) and H$\gamma$ (from
the \espa\ spectropolarimeter). We discarded the other hydrogen
lines observed with \espa\ because of the uncertainties in the continuum
normalisation. In the temperature range expected for WASP-12, hydrogen lines
are extremely sensitive to temperature variations and insensitive to \logg\ 
variations, and are therefore good temperature indicators. We found
\Teff=6250$\pm$100\,K, in good agreement with \citet{hebb}. The
uncertainty estimate considered both the quality of the observations and the
uncertainties in the normalisation.
Figure~\ref{hydrogen} shows the comparison between the observed H$\alpha$ line
profile and the synthetic profiles calculated with the adopted \Teff\ of
6250$\pm$100\,K. The poor fit of the hydrogen line core is due to the adopted
LTE approximation \citep[see e.g.][]{mashonkina09}.

\begin{figure}
\begin{center}
\includegraphics[width=\hsize,clip]{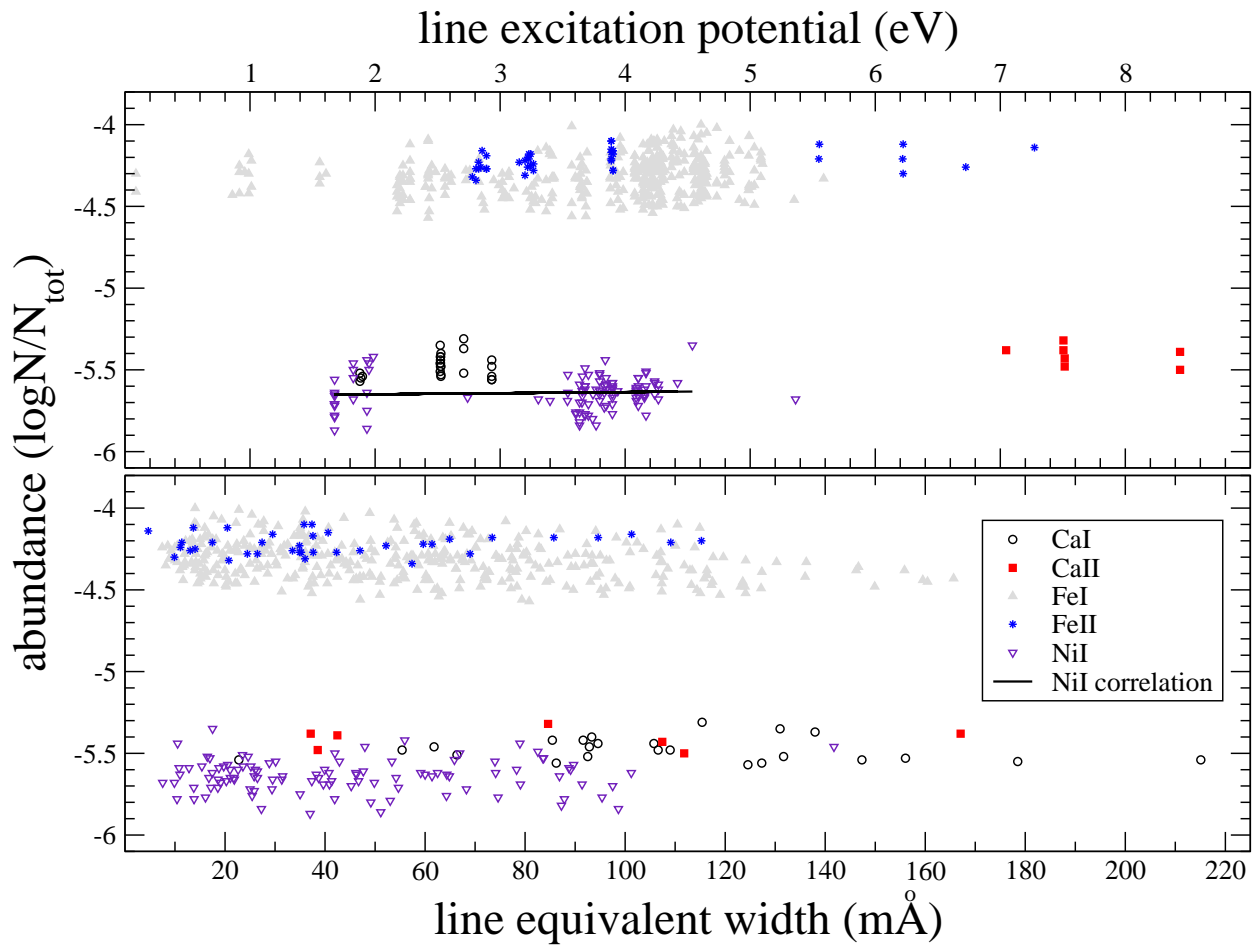}
\caption{Top panel: abundance obtained for each measured line of \ion{Ca}{1}
(open circles), \ion{Ca}{2} (filled squares), \ion{Fe}{1} (filled triangles),
\ion{Fe}{2} (stars), and \ion{Ni}{1} (open downside triangles) as a
function of the line \exc. The black full line shows, as an
example, the abundance-\exc\ correlation for \ion{Ni}{1}: 0.00637$\pm$0.01270,
consistent with zero. Bottom panel: abundance obtained for
each measured line of the ions given in the top panel, as a function of the
measured line equivalent width. The effective temperature from metallic lines
is derived eliminating the correlation shown in the top panel, while \vmic\ 
is determined eliminating the correlation shown in the bottom panel.}
\label{equilibria}
\end{center}
\end{figure}

Another spectroscopic indicator for \Teff\ is given by the analysis of
metallic lines, that we performed on the \espa\ spectrum. In particular,
\Teff\ is determined by eliminating the correlation between abundance and
excitation potential (\exc) for the selected lines of a given ion/element.
The top panel of Fig.~\ref{equilibria} shows the correlation between abundance
and \exc\ for all measured lines of \ion{Ca}{1}, \ion{Ca}{2}, \ion{Fe}{1},
\ion{Fe}{2}, and \ion{Ni}{1}. Figure~\ref{equilibria}, produced using the final
adopted fundamental parameters, shows no significant correlation between
abundance and \exc\ for all ions, except \ion{Fe}{1}, for which we
registered a slightly positive correlation (0.02263$\pm$0.01139), that would
be eliminated by a higher \Teff. What we found here for \ion{Fe}{1} resambles
what remarked by \citet{hd49933} for HD~49933 and HD~32115 (both stars have a
\Teff\ slightly higher than WASP-12): an effective temperature determination
based only on the analysis of \ion{Fe}{1} lines leads systematically to a
\Teff\ that is substantially higher (by about 5\%), compared to the one
obtained with other ions and in particular with other temperature indicators,
such as hydrogen lines \footnote{We do not know the precise origin of this
phenomenon and also whether it is present in a large temperature range or
just for F-type stars. We believe that more work should be done in this
respect, in particular because the \Teff\ determination based on the
abundance-\exc\ equilibrium is widely used to determine the effective
temperature of stars in a broad \Teff\ range.}. For this reason, we
decided to use the analysis of the metallic lines only as consistency check
of \Teff\ derived from the hydrogen lines. In addition to the ions shown in
Fig.~\ref{equilibria}, we included in the consistency check also \ion{C}{1},
\ion{Si}{1}, \ion{Sc}{2}, \ion{Ti}{1}, \ion{Ti}{2}, \ion{V}{1}, \ion{Cr}{1},
\ion{Cr}{2}, \ion{Mn}{1}, \ion{Co}{1}, and \ion{Y}{2}, obtaining the requested
equilibrium for all of them. The number of lines adopted to measure the
abundance-\exc\ correlation for each ion is the same as the one we used to
derive the final ion abundance, and it is listed in Table~\ref{abundance}.

\begin{figure}
\begin{center}
\includegraphics[width=\hsize,clip]{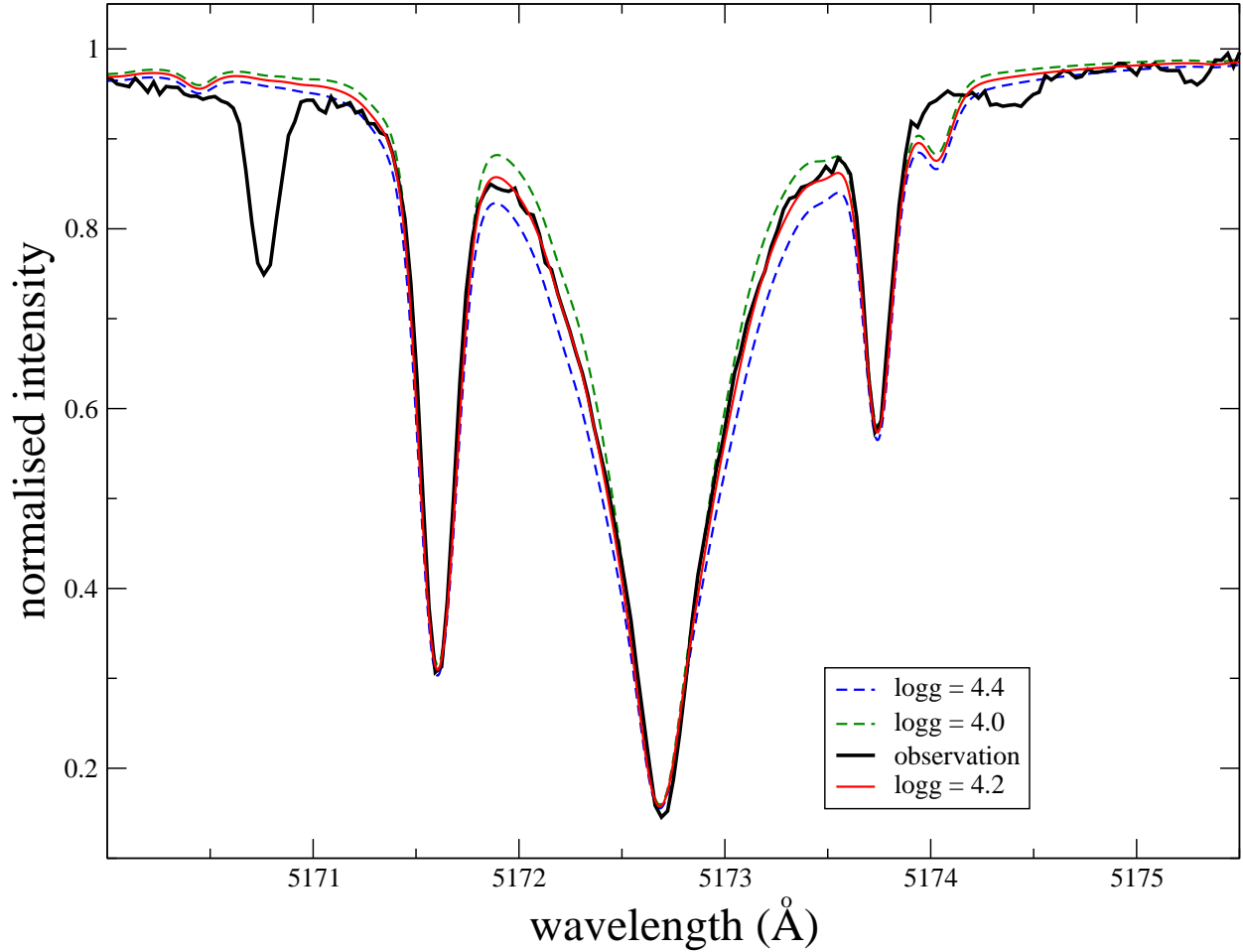}
\caption{Comparison between the observed profile of the 5172\,\AA\ \ion{Mg}{1}
line (black solid line) and synthetic profiles calculated with the final
adopted surface gravity of \logg=4.2$\pm$0.2. We adopted the same combination
of \loggf\ and Van der Waals damping constants as in \citet{hd49933}:
oscillator strengths from \citet{Aldenius} and damping constants from
\citet{fuhrmann}.}
\label{mglines}
\end{center}
\end{figure}

The surface gravity was derived from two independent methods based on ({\it i})
line profile fitting of gravity-sensitive metal lines with developed wings
and ({\it ii}) ionisation balance for several elements. The first method is
described in \citet{fuhrmann} and uses the fact that the wings of the
\ion{Mg}{1} lines at $\lambda$\,5167, 5172 and 5183\,\AA\ are very sensitive
to \logg\ variations. In practice we first derived the Mg abundance from other
\ion{Mg}{1} lines without developed wings, such as $\lambda$\,5711 and
5785\,\AA, and then we fit the wings of the gravity indicator lines by
tuning the \logg\ value. To apply this method, very accurate \loggf\ values
and Van der Waals (log\,$\mathbf{\gamma_{\rm Waals}}$) damping constants are
required for all the lines. We adopted the set of line parameters used by
\citet{hd49933} and included the uncertainty in these parameters in the
uncertainty in \logg. We obtained a \logg\ value of 4.2$\pm$0.2, in good
agreement with \citet{hebb}. Our line profile fit of the \ion{Mg}{1} lines
with developed wings is shown in Fig.~\ref{mglines}.

The second method, the ionisation equilibrium, is often used to derive the
surface gravity, but this method is extremely sensitive to the \nlte\ effects
present for each ion/element, while Mg lines with developed wings (less
sensitive to \nlte\ effects) are more suitable as \logg\ indicators
\citep{zhao}. For this reason we decided to keep the Mg lines as our primary
\logg\ indicator, and checked the result with the ionisation
equilibrium. Adopting the \logg\ value obtained from the analysis of the
\ion{Mg}{1} lines with developed wings and taking into account the abundance
uncertainties, we satisfied the ionisation equilibrium for every element with
two analysed ions.

Our main source for the atomic parameters of spectral lines is the \vald\ 
database with the default configuration. LTE abundance analysis was based on
equivalent widths, analysed with a modified version \citep{vadim} of the
WIDTH9 code \citep{kurucz1993}. For all analysed elements/ions we used almost
all unblended spectral lines with accurate atomic parameters available in the
wavelength range 4240--9900\,\AA, except lines in spectral regions where the
continuum normalisation was too uncertain. For blended lines, lines subjected
to hyperfine splitting (\hfs), lines situated in the wings of the hydrogen
lines or for very shallow lines of specific ions we derived the line abundance
by performing synthetic spectrum calculations with the \synth\ code. The \hfs\ 
constants for abundance calculations were taken from \citet{LWHS} for
\ion{Eu}{2} lines and from \citet{SLN} for the \ion{Li}{1} line at
$\lambda$\,6707\,\AA. For the Barium abundance we used the \ion{Ba}{2} lines
at $\lambda$\,5853.7\,\AA\ and 6496.9\,\AA\ for which we do not expect any
relevant \hfs\ effect \citep{mashonkina06}. A line-by-line abundance list
with the equivalent width measurements, adopted oscillator strengths, and
their sources is given in Table~\ref{linelist} (see the online material for
the complete version of the table).
\begin{table}
\caption[ ]{Linelist of the lines used for the abundance analysis.
The first and second columns list respectively the \exc\ (in eV) 
and the \loggf\ value for each line. Columns five and six list the 
equivalent width in m\AA\ and the abundance for each line, while 
the last column gives the reference for the \loggf\ value. 
Spectral lines for which the abundance was measured with 
synthetic spectra, instead of equivalent widths, present an ''S" 
instead of the equivalent width value. Lines marked with ''*" 
are subject to hyperfine structure, discussed in detail in 
the main text, while lines marked with ''\#" are multiplets (doublets
or triplets) and in these cases we listed only the strongest line.
The entire table can be viewed in the electronic version of the Journal.}
\label{linelist}
\begin{center}
\begin{tabular}{c|c|c|r|r|r}
\hline
\hline
Ion & & & & & \\
Wavelength & $\chi_{\mathrm{excit}}$ & log $gf$ & EQW	 & abundance & Ref. $\log gf$ \\
$\AA$	   & eV 		     &  	& m$\AA$ & dex       &  	      \\
\hline                                                                                                    
\ion{Li}{1} & & & & & \\
6707.7610* &   0.000  &  -0.009 &       S &   -9.47 & SLN \\
\hline
\ion{C}{1} & & & & & \\
5023.8389  &   7.946  &  -2.209 &   12.14 &   -3.49 & WFD \\
5800.6016  &   7.946  &  -2.338 &    9.51 &   -3.43 & WFD \\
6014.8300  &   8.643  &  -1.585 &   15.74 &   -3.40 & WFD \\
6671.8450  &   8.851  &  -1.651 &    9.97 &   -3.36 & WFD \\
7111.4694  &   8.640  &  -1.086 &   26.93 &   -3.55 & WFD \\
7116.9879  &   8.647  &  -0.907 &   36.47 &   -3.52 & WFD \\
...        &   ...    &  ...    &   ...   &   ...   & ... \\
\hline
\end{tabular}
\end{center}
\smallskip 

SLN - \citet{SLN};\\
WFD - \citet{WFD};\\
...
\end{table}

The microturbulence velocity was determined by minimising the correlation
between equivalent width and abundance for several ions, as shown in
Fig.~\ref{equilibria} for \ion{Ca}{1}, \ion{Ca}{2}, \ion{Fe}{1},
\ion{Fe}{2}, and \ion{Ni}{1}. To employ this method we used all ions for
which the measured spectral lines covered a large range in equivalent width.
In particular we took into account simultaneously: \ion{Si}{1}, \ion{Ti}{1},
\ion{Ti}{2}, \ion{Cr}{1}, \ion{Cr}{2}, \ion{Mn}{1}, and \ion{Y}{2}, in addition
to the ones present in Fig.~\ref{equilibria}. The final adopted
\vmic\ is 1.2$\pm$0.3\,\kms. The given error bar is the range of values
resulting from minimisation of the correlation for each ion we considered.

The projected rotational velocity and macroturbulence (\vmac) were determined
by fitting synthetic spectra of several carefully selected lines to the
observed spectrum. Given the \vsini-\vmac\ degeneracy, we followed
\citet{valenti}: we measured \vmac\ assuming \vsini=0\,\kms\ and then \vsini\ 
assuming \vmac=4.75\,\kms\
\citep[from the \Teff-\vmac\ relation published by][]{valenti}.
In the first case we obtained \vmac=7.0$\pm$0.6\,\kms, while in the second
case we obtained \vsini=4.6$\pm$0.5\,\kms. In conclusion \vsini\ is in the
range 0--4.6$\pm$0.5\,\kms, while \vmac\ lays between 4.75 and
7.0$\pm$0.6\,\kms. Only with a careful analysis of the Rossiter-McLaughlin
(RM) effect will it be possible to precisely measure \vsini.

\begin{figure}
\begin{center}
\includegraphics[width=\hsize,clip]{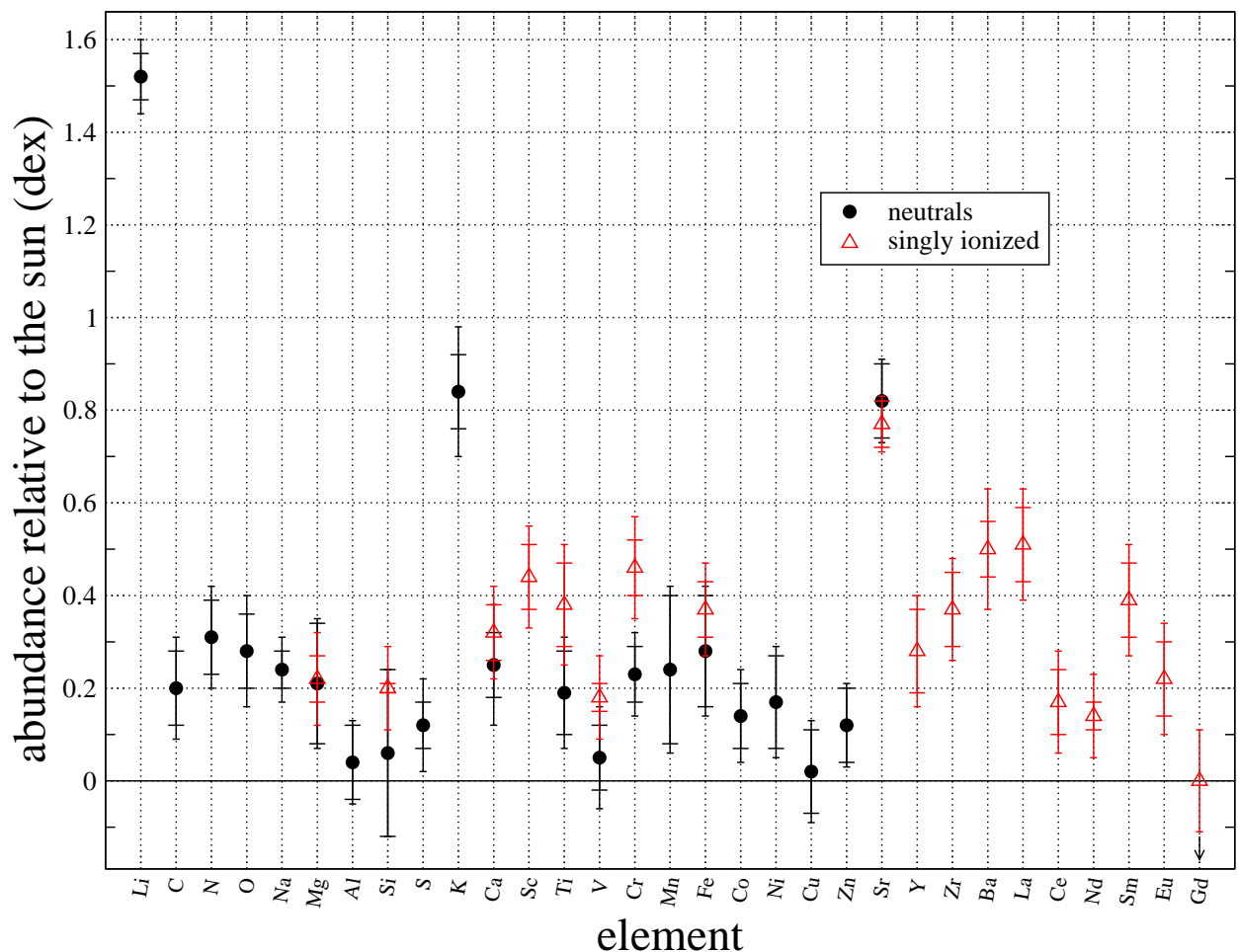}
\caption{Ion abundance relative to the Sun \citep{met05} of the WASP-12
atmosphere. Full points show the abundance of the neutral elements,
while the open triangles show the abundance of the singly ionised
elements. Each abundance value is shown with two uncertainty values: the
standard deviation from the mean (column 3 of Table~\ref{error}) and the
total uncertainty (column eight of Table~\ref{error}). We believe the real
uncertainty lies between these two values.}
\label{fig.abn}
\end{center}
\end{figure}

The final WASP-12 abundances, in $\log (N/N_{\rm tot})$, are given in
Table~\ref{abundance} and the atmospheric abundance pattern is shown in
Fig.~\ref{fig.abn} in comparison to the solar abundances \citep{met05}.
While the large overabundance of K would disappear if \nlte\ effects
are taken into account \citep{takeda96}, the Sr overabundance is
genuine since \nlte\ effects are expected to be less then 0.05\,dex for the
Sr lines we analysed \citep[][and references therein]{mashonkina07}.
In general we expect small \nlte\ effects for WASP-12 due to the high stellar
metallicity.
\begin{table}
\caption[ ]{LTE atmospheric abundances of WASP-12 with error bar 
estimates based on the internal scatter from the number of analysed 
lines, $n$. The fourth column gives the WASP-12 abundances in dex
relative to the solar values from \citet{met05}. The last 
column gives the abundances of the solar atmosphere from \citet{met05}. 
The Lithium and Europium abundances take hyperfine structure in the lines into
account. The \ion{Gd}{2} abundance is an upper limit.
The symbol \# indicates the ions for which the abundance 
was derived from line profile fitting, instead of equivalent widths.}
\tiny
\label{abundance}
\begin{center}
\begin{tabular}{l|cc|c|c}
\hline
\hline
Ion &\multicolumn{3}{|c|}{WASP-12}& Sun \\                    
    &$\log (N/N_{\rm tot})$ & $n$ &[$N_{\rm el}/N_{\rm tot}$] & $\log (N/N_{\rm tot})$  \\
\hline                                                                                                    
LiI \# & ~~$-$9.47$\pm$0.05 &  2 & ~$+$1.52 & ~$-$10.99~ \\	 
CI     & ~~$-$3.45$\pm$0.08 & 14 & ~$+$0.20 & ~~$-$3.65~ \\	 
NI  \# & ~~$-$3.95	    &  1 & ~$+$0.31 & ~~$-$4.26~ \\	 
OI     & ~~$-$3.10	    &  1 & ~$+$0.28 & ~~$-$3.38~ \\	 
NaI    & ~~$-$5.63$\pm$0.04 &  4 & ~$+$0.24 & ~~$-$5.87~ \\	 
MgI    & ~~$-$4.30$\pm$0.13 &  4 & ~$+$0.21 & ~~$-$4.51~ \\	 
MgII   & ~~$-$4.29$\pm$0.05 &  3 & ~$+$0.22 & ~~$-$4.51~ \\	 
AlI    & ~~$-$5.63$\pm$0.08 &  4 & ~$+$0.04 & ~~$-$5.67~ \\	 
SiI    & ~~$-$4.47$\pm$0.18 & 60 & ~$+$0.06 & ~~$-$4.53~ \\	 
SiII   & ~~$-$4.33$\pm$0.01 &  2 & ~$+$0.20 & ~~$-$4.53~ \\	 
SI  \# & ~~$-$4.78$\pm$0.05 &  8 & ~$+$0.12 & ~~$-$4.90~ \\	 
KI     & ~~$-$6.12	    &  1 & ~$+$0.84 & ~~$-$6.96~ \\	 
CaI    & ~~$-$5.48$\pm$0.07 & 24 & ~$+$0.25 & ~~$-$5.73~ \\	 
CaII   & ~~$-$5.41$\pm$0.06 &  7 & ~$+$0.32 & ~~$-$5.73~ \\	 
ScII   & ~~$-$8.55$\pm$0.07 &  8 & ~$+$0.44 & ~~$-$8.99~ \\	 
TiI    & ~~$-$6.95$\pm$0.09 & 37 & ~$+$0.19 & ~~$-$7.14~ \\	 
TiII   & ~~$-$6.76$\pm$0.09 & 24 & ~$+$0.38 & ~~$-$7.14~ \\	 
VI     & ~~$-$7.99$\pm$0.07 & 11 & ~$+$0.05 & ~~$-$8.04~ \\	 
VII    & ~~$-$7.86$\pm$0.03 &  2 & ~$+$0.18 & ~~$-$8.04~ \\	 
CrI    & ~~$-$6.17$\pm$0.06 & 32 & ~$+$0.23 & ~~$-$6.40~ \\	 
CrII   & ~~$-$5.94$\pm$0.06 & 12 & ~$+$0.46 & ~~$-$6.40~ \\	 
MnI    & ~~$-$6.41$\pm$0.16 & 16 & ~$+$0.24 & ~~$-$6.65~ \\	 
FeI    & ~~$-$4.31$\pm$0.12 &389 & ~$+$0.28 & ~~$-$4.59~ \\	 
FeII   & ~~$-$4.22$\pm$0.06 & 38 & ~$+$0.37 & ~~$-$4.59~ \\	 
CoI    & ~~$-$6.98$\pm$0.07 & 12 & ~$+$0.14 & ~~$-$7.12~ \\	 
NiI    & ~~$-$5.64$\pm$0.10 &105 & ~$+$0.17 & ~~$-$5.81~ \\	 
CuI    & ~~$-$7.81$\pm$0.09 &  4 & ~$+$0.02 & ~~$-$7.83~ \\	 
ZnI    & ~~$-$7.32	    &  1 & ~$+$0.12 & ~~$-$7.44~ \\	 
SrI \# & ~~$-$8.30	    &  1 & ~$+$0.82 & ~~$-$9.12~ \\	 
SrII\# & ~~$-$8.35$\pm$0.05 &  3 & ~$+$0.77 & ~~$-$9.12~ \\	 
YII    & ~~$-$9.55$\pm$0.09 & 10 & ~$+$0.28 & ~~$-$9.83~ \\	 
ZrII   & ~~$-$9.08	    &  1 & ~$+$0.37 & ~~$-$9.45~ \\	 
BaII   & ~~$-$9.37$\pm$0.06 &  2 & ~$+$0.50 & ~~$-$9.87~ \\	 
LaII   & ~$-$10.40	    &  1 & ~$+$0.51 & ~$-$10.91~ \\	 
CeII   & ~$-$10.29$\pm$0.07 &  3 & ~$+$0.17 & ~$-$10.46~ \\	 
NdII   & ~$-$10.45$\pm$0.03 &  4 & ~$+$0.14 & ~$-$10.59~ \\	 
SmII   & ~$-$10.64	    &  1 & ~$+$0.39 & ~$-$11.03~ \\	 
EuII\# & ~$-$11.30	    &  1 & ~$+$0.22 & ~$-$11.52~ \\	 
GdII\# & ~$\leq-$10.92      &  1 & ~$+$0.00 & ~$-$10.92~ \\	 
\hline											      
\Teff     &\multicolumn{3}{|c|}{6250~K}    & 5777~K  \\ 			    
\logg     &\multicolumn{3}{|c|}{4.20~~~}   & 4.44~~~~\\ 			     
\hline											  
\end{tabular}
\end{center}
\end{table}

The stellar metallicity ($Z$) is defined as follows:
\begin{equation}
\label{Z}
Z_{\rm star}=\frac{\sum_{a \geq 3}m_{a}10^{\log(N_{a}/N_{tot})}}{\sum_{a \geq 1}m_{a}10^{\log(N_{a}/N_{tot})}},
\end{equation}
where $a$ is the atomic number of an element with atomic mass $m_{\rm a}$.
Our abundances imply a metallicity of $Z$=0.021$\pm$0.002\,dex, adopting
the solar abundances by \citet{met05} for all the elements that we did not
analyse.

The $Z$ value adopted to characterise isochrones is calculated with the
following approximation:
\begin{equation}
\label{isoZ}
Z_{\mathrm star} \simeq 10^{([Fe/H]_{\mathrm star}-[Fe/H]_{\odot})} \cdot Z_{\odot}\ ,
\end{equation}
assuming $Z_{\odot}$=0.019\,dex. We recalculated the $Z$ of WASP-12
according to this approximation obtaining $Z$=0.036$\pm$0.002\,dex.
\subsection{Abundance uncertainties}
The abundance uncertainties for each ion, shown in Table~\ref{abundance}, are
the standard deviations of the mean abundance obtained from the individual
line abundances. Following \citet{fossati2009}, it is possible to conclude
that in case of ions with a sufficiently high number of lines, the internal
scatter for each ion includes the uncertainties due to equivalent width
measurement and continuum normalisation. In addition, from plotting the
abundance scatter as a function of the number of lines, we can also infer an
internal uncertainty of 0.08\,dex, that can be applied as mean scatter when
only one line of a certain ion is measured.

Table~\ref{error} shows the variation in abundance for each analysed ion,
caused by the change of one fundamental parameter by $+1\sigma$, keeping
fixed the other parameters.
\begin{table*}
\caption{
Uncertainty sources for the abundances of WASP-12.
The third column shows the standard deviation $\sigma_{\rm abn}$ (scatt.) 
of the mean abundance obtained from different spectral lines (internal 
scattering); a blank means that only a single line was used, 
and we estimated the internal scattering to be 0.08\,dex. Note that these 
values are identical to those given in Table~\ref{abundance}. Columns 4, 5, 
and 6 give the variation in abundance estimated by increasing
\Teff\ by 100\,K, \logg\ by 0.2\,dex, and \vmic\ by 0.3\,\kms,
respectively. Column~7 gives the the mean error bar calculated adding 
the systematic uncertainties given in columns 4, 5 and 6 in quadrature i.e.,
$\sigma_{\rm abn}^2$\,(syst.)    = 
$\sigma_{\rm abn}^2$\,(\Teff)  + 
$\sigma_{\rm abn}^2$\,(\logg)  + 
$\sigma_{\rm abn}^2$\,(\vmic).
Column~8 gives the total mean error bar: 
$\sigma_{\rm abn}^2$\,(tot.)    =
$\sigma_{\rm abn}^2$\,(syst.)  +
$\sigma_{\rm abn}^2$\,(scatt.).
}
\tiny
\label{error}
\begin{center}
\begin{tabular}{lrrrrrrr}
\hline
\hline
\multicolumn{1}{c}{Ion         }        &  
\multicolumn{1}{c}{abundance   }        &  
\multicolumn{1}{c}{$\sigma_{\rm abn}$\,(scatt.)}  &  
\multicolumn{1}{c}{$\sigma_{\rm abn}$\,(\Teff) }  &  
\multicolumn{1}{c}{$\sigma_{\rm abn}$\,(\logg) }  &  
\multicolumn{1}{c}{$\sigma_{\rm abn}$\,(\vmic) }  &  
\multicolumn{1}{c}{$\sigma_{\rm abn}$\,(syst.) }  &  
\multicolumn{1}{c}{$\sigma_{\rm abn}$\,(tot.)  }  \\ 
\multicolumn{1}{c}{            }        & 
\multicolumn{1}{c}{$\log (N/N_{\rm tot})$} & 
\multicolumn{1}{c}{(dex)       }        & 
\multicolumn{1}{c}{(dex)       }        & 
\multicolumn{1}{c}{(dex)       }        & 
\multicolumn{1}{c}{(dex)       }        & 
\multicolumn{1}{c}{(dex)       }        & 
\multicolumn{1}{c}{(dex)       }        \\ 
\hline
LiI  & ~$-$9.47 & 0.05 &    0.06 & $-$0.01 &    0.00 & 0.06 & 0.08 \\   
CI   & ~$-$3.45 & 0.08 & $-$0.05 &    0.05 &    0.00 & 0.07 & 0.11 \\   
NI   & ~$-$3.95 &      & $-$0.06 &    0.05 &    0.00 & 0.08 & 0.11 \\   
OI   & ~$-$3.10 &      & $-$0.08 &    0.05 &    0.00 & 0.09 & 0.12 \\   
NaI  & ~$-$5.63 & 0.04 &    0.04 & $-$0.04 & $-$0.02 & 0.06 & 0.07 \\   
MgI  & ~$-$4.30 & 0.13 &    0.04 & $-$0.02 & $-$0.02 & 0.05 & 0.14 \\   
MgII & ~$-$4.29 & 0.05 & $-$0.08 &    0.04 & $-$0.02 & 0.09 & 0.10 \\   
AlI  & ~$-$5.63 & 0.08 &    0.03 & $-$0.01 & $-$0.01 & 0.03 & 0.09 \\   
SiI  & ~$-$4.47 & 0.18 &    0.02 & $-$0.01 & $-$0.01 & 0.02 & 0.18 \\   
SiII & ~$-$4.33 & 0.01 & $-$0.07 &    0.06 & $-$0.02 & 0.09 & 0.09 \\   
SI   & ~$-$4.78 & 0.05 &    0.06 & $-$0.06 &    0.00 & 0.08 & 0.10 \\   
KI   & ~$-$6.12 &      &    0.08 & $-$0.08 & $-$0.04 & 0.12 & 0.14 \\   
CaI  & ~$-$5.48 & 0.07 &    0.07 & $-$0.05 & $-$0.06 & 0.10 & 0.13 \\   
CaII & ~$-$5.41 & 0.06 & $-$0.06 &    0.04 & $-$0.02 & 0.07 & 0.10 \\   
ScII & ~$-$8.55 & 0.07 &    0.01 &    0.07 & $-$0.04 & 0.08 & 0.11 \\   
TiI  & ~$-$6.95 & 0.09 &    0.08 &    0.00 & $-$0.02 & 0.08 & 0.12 \\   
TiII & ~$-$6.76 & 0.09 &    0.00 &    0.07 & $-$0.06 & 0.09 & 0.13 \\   
VI   & ~$-$7.99 & 0.07 &    0.09 &    0.00 & $-$0.01 & 0.09 & 0.11 \\   
VII  & ~$-$7.86 & 0.03 &    0.00 &    0.08 & $-$0.01 & 0.08 & 0.09 \\   
CrI  & ~$-$6.17 & 0.06 &    0.06 & $-$0.01 & $-$0.02 & 0.06 & 0.09 \\   
CrII & ~$-$5.94 & 0.06 & $-$0.02 &    0.08 & $-$0.03 & 0.09 & 0.11 \\   
MnI  & ~$-$6.41 & 0.16 &    0.07 &    0.00 & $-$0.03 & 0.08 & 0.18 \\   
FeI  & ~$-$4.31 & 0.12 &    0.06 & $-$0.02 & $-$0.04 & 0.07 & 0.14 \\   
FeII & ~$-$4.22 & 0.06 & $-$0.02 &    0.07 & $-$0.04 & 0.08 & 0.10 \\   
CoI  & ~$-$6.98 & 0.07 &    0.07 &    0.00 & $-$0.01 & 0.07 & 0.10 \\   
NiI  & ~$-$5.64 & 0.10 &    0.06 & $-$0.01 & $-$0.03 & 0.07 & 0.12 \\   
CuI  & ~$-$7.81 & 0.09 &    0.06 &    0.00 & $-$0.02 & 0.06 & 0.11 \\   
ZnI  & ~$-$7.32 &      &    0.02 &    0.02 & $-$0.02 & 0.03 & 0.09 \\   
SrI  & ~$-$8.30 &      &    0.02 & $-$0.01 & $-$0.03 & 0.04 & 0.09 \\   
SrII & ~$-$8.35 & 0.05 &    0.00 &    0.02 & $-$0.02 & 0.03 & 0.06 \\   
YII  & ~$-$9.55 & 0.09 &    0.01 &    0.08 & $-$0.03 & 0.09 & 0.12 \\   
ZrII & ~$-$9.08 &      &    0.01 &    0.08 & $-$0.01 & 0.08 & 0.11 \\   
BaII & ~$-$9.37 & 0.06 &    0.04 &    0.02 & $-$0.11 & 0.12 & 0.13 \\   
LaII & $-$10.40 &      &    0.03 &    0.08 & $-$0.01 & 0.09 & 0.12 \\   
CeII & $-$10.29 & 0.07 &    0.03 &    0.08 & $-$0.01 & 0.09 & 0.11 \\   
NdII & $-$10.45 & 0.03 &    0.03 &    0.08 & $-$0.01 & 0.09 & 0.09 \\   
SmII & $-$10.64 &      &    0.02 &    0.08 & $-$0.02 & 0.08 & 0.12 \\   
EuII & $-$11.30 &      &    0.03 &    0.08 & $-$0.01 & 0.09 & 0.12 \\   
GdII & $\leq-$10.92&   &    0.02 &    0.08 &    0.00 & 0.08 & 0.11 \\   
\hline								
\end{tabular}
\end{center}
\end{table*}

Table~\ref{error} shows that the main source of uncertainty varies according to
the element/ion (e.g. for the Fe-peak elements, neutrals are more sensitive to
temperature variations, while ions are more sensitive to \logg\ variations)
and in some cases to the selected lines (e.g. the two \ion{Ba}{2} lines
selected to measure the Ba abundance are rather strong, therefore the Ba
abundance is strongly dependent on \vmic\ variations).

Assuming the different uncertainties in the abundance determination are
independent (though actually the systematic uncertainties will be correlated),
we derived a pessimistic final error bar using standard error propagation
theory, given in columns seven and eight of Table~\ref{error}.
Using the propagation theory we considered the situation where the
determination of each fundamental parameter is an independent process. The
mean value of the LTE uncertainties given in column eight of
Table~\ref{error} is 0.11\,dex. Due to the fact that for the parameter
determination of both \Teff, \logg\ and \vmic\ we took into account all
possible systematics (except \nlte), we believe that the abundance
uncertainties given in the last column of Table~\ref{error} can be considered
as upper limits and that the real error bars lie between the values given in
column three and eight of Table~\ref{error}.
\section{High precision magnetic field search}\label{mag.field}
One of the main goals of our analysis is to search for a global stellar
magnetic field in WASP-12. The \espa\ spectropolarimeter yields high
resolution and high SNR spectra of both Stokes $I$ and $V$ allowing this
search. To detect the presence of a global magnetic field and measure its
strength we used the Least-Squares Deconvolution technique (hereafter LSD),
adopting a code written by one of us (O.~Kochukhov).

LSD is a cross-correlation technique developed for the detection and
measurement of weak polarisation signatures in stellar spectral lines. The
method is described in detail by \citet{donatietal1997} and \citet{wade2000}.
We decided to use the LSD approach to detect a magnetic field in WASP-12
since this method is the most precise currently available, especially
for stars with rich line spectra and low projected rotational velocity
(\vsini), such as WASP-12.

\begin{figure}
\begin{center}
\includegraphics[width=17cm]{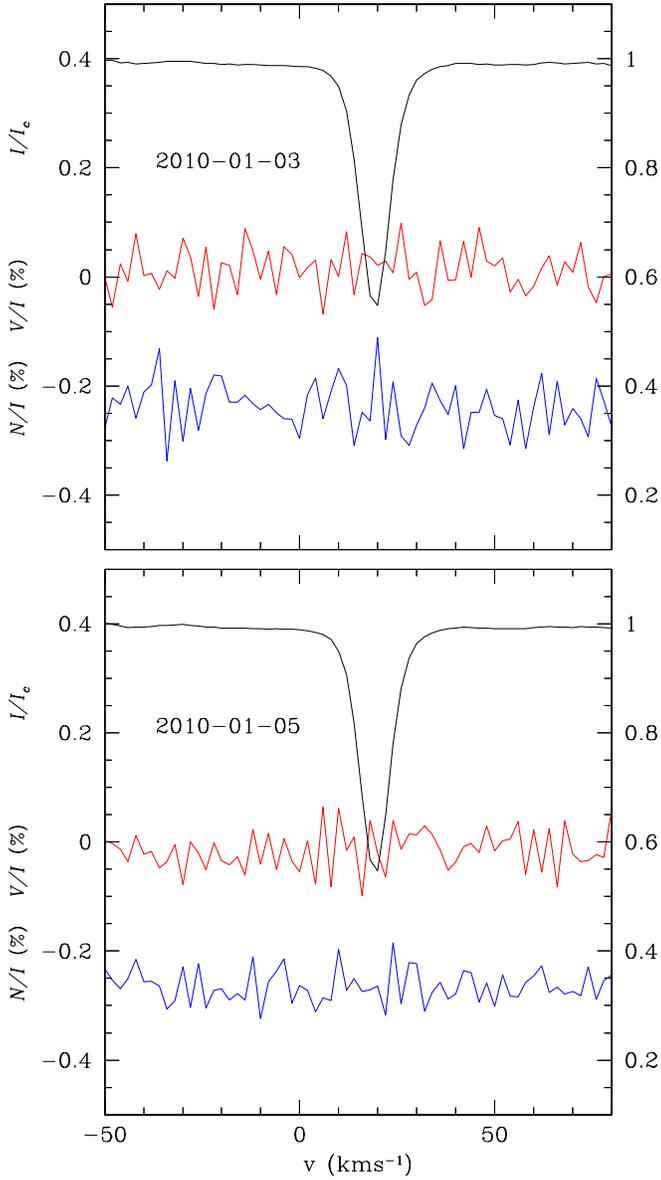}
\caption{LSD profiles from the spectra obtained on January 3rd (top panel),
corresponding to an orbital phase of 0.80, and January 5th 2010 (bottom panel),
corresponding to an orbital phase of 0.81. In each panel, the top black
line shows Stokes $I$, normalised to the continuum (the corresponding
units are given in the right $y$-axis). The middle red line corresponds to
the $V/I$ profile and the bottom blue line to the null profile
\citep{donatietal1997}; their corresponding units are given in
the left $y$-axis. The Stokes $I$ LSD profile is centered on the stellar
radial velocity and the null profile is shifted downward by an arbitrary
offset.}
\label{lsd}
\end{center}
\end{figure}

We applied the LSD technique to the Stokes $V$ spectra from each of the two
nights using about 6\,450 atomic spectral lines with the only cut-off
criterion based on the calculated line depth ($>$10\%), and in both cases
no magnetic field was found. From the spectrum of January 3rd we obtained
$\langle B_z\rangle=2.3\pm5.3$\,G, with a SNR of the Stokes $V$ LSD profile
of 2400, while from the second night spectrum we obtained
$\langle B_z\rangle=-10.1\pm4.2$\,G, with a SNR of the Stokes $V$ LSD profile
of 3380. We obtained similar values also from both null profiles.
Figure~\ref{lsd} shows the LSD profiles.

\begin{figure}
\begin{center}
\includegraphics[width=\hsize,clip]{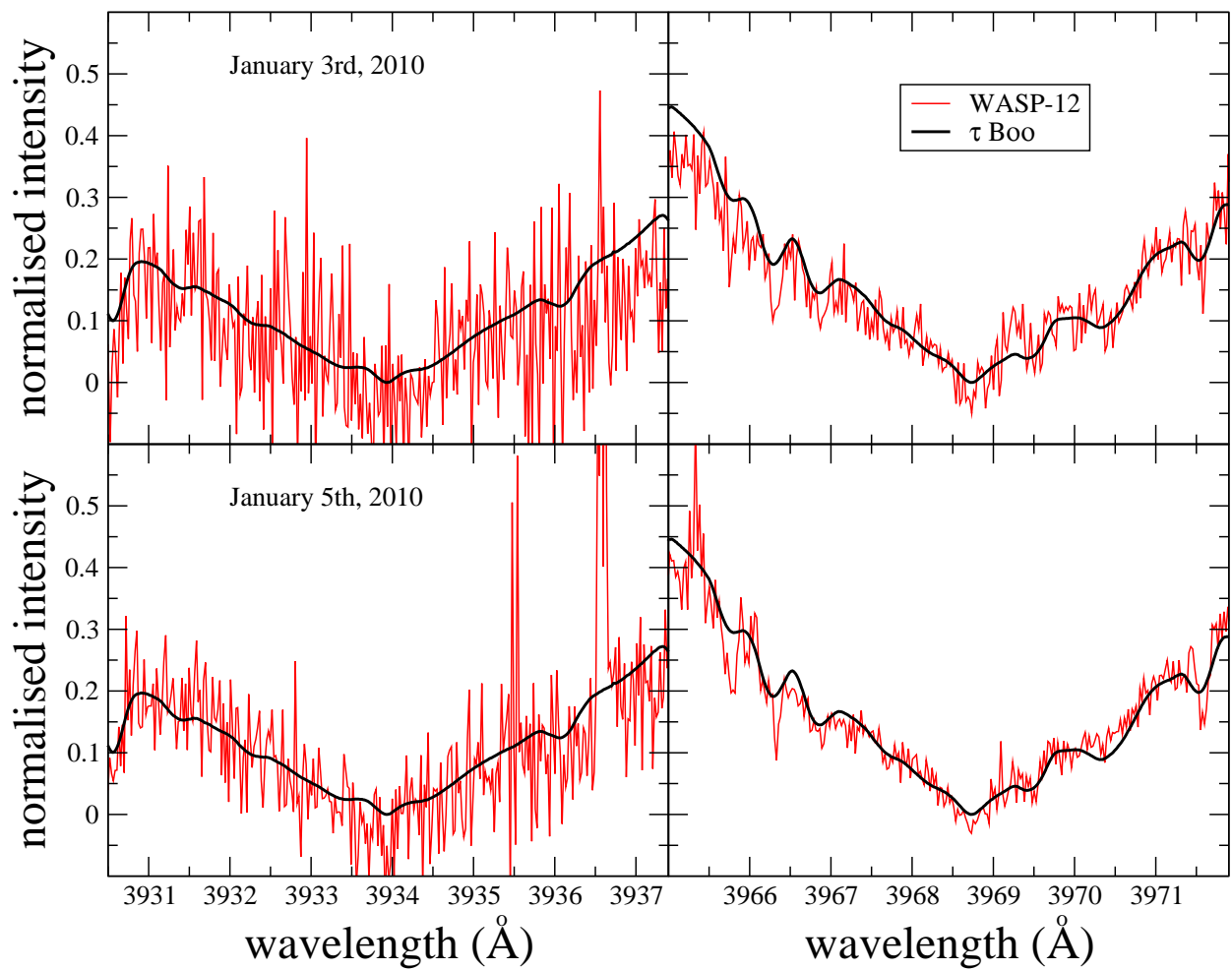}
\caption{Comparison of the observed Ca K and H line (left and right panels,
respectively) profiles between WASP-12 (thin line) and the planet hosting
star $\tau$\,Boo \citep[thick line,][]{es05}. The two upper panels show the
WASP-12 spectrum obtained on January 3rd, 2010, while the two lower panels
show the spectrum obtained on January 5th, 2010.}
\label{caHK}
\end{center}
\end{figure}

In late-type stars, the stellar magnetic field is directly connected to the
chromospheric activity, that can be monitored with the \ion{Ca}{2} H and K
lines. In Fig.~\ref{caHK} we compare the profiles of the \ion{Ca}{2} H and K
lines, observed with \espa, with the mean line profiles of $\tau$\,Boo
\citep{es05} obtained averaging several CFHT spectra acquired with the GECKO
spectrograph\footnote{{\it http://www.cfht.hawaii.edu/Instruments/Spectroscopy/Gecko/}}.
This comparison is particularly valuable because both stars are planet
hosting and have similar fundamental parameters and age \citep{gonzalez2010a},
where the difference is mainly in the \vsini\ values ($\tau$\,Boo has a
\vsini\ of about 13.5\,\kms). Figure~\ref{caHK} does not show the presence
of any anomaly in WASP-12 \ion{Ca}{2} H and K line cores. \citet{knutson}
determined the $\log (R'_{HK})$ chromospheric stellar acitvity parameter in
a set of planet hosting stars, reporting for WASP-12 the remarkably low value
of -5.500, the lowest in their sample.

The bottom panel of Figure~1 in \citet{fossati2010} shows the core of the
\ion{Mg}{2} UV resonance lines (further stellar activity indicators) of
WASP-12 and here the lack of any line core emission is clear, in agreement
with the low level of stellar activity reported by \citet{knutson}.

It is well known that the chromospheric activity is strongly correlated with
the stellar rotational velocity, that for WASP-12 is unknown. Given the very
small RM effect shown by \citet{husnoo}, it is probable that the
low stellar activity of WASP-12 might be connected to a low rotational
velocity. It is also possible that the star is either passing through a
period of low activity, or WASP-12 is much older than reported by
\citet{hebb} (more than 1\,Gyr older), or the typical signs of stellar
activity, such as emission in the cores of the \ion{Mg}{2} resonance lines,
are absorbed by the material lost by the planet and falling onto the star
\citep[see][]{li2010,fossati2010}. A more thorough analysis and discussion
of the stellar activity and of the reasons behind its low level will be
given in a following work, now in preparation.
\section{Discussion}\label{discussion}
\subsection{Atmospheric parameters and convection}
Because WASP-12 has a relatively low effective temperature, its hydrogen lines
are ideal indicators of atmospheric T-P structure. Generally, fitting
hydrogen line profiles, rather than any other atomic line, provides an
accurate estimate of \Teff. On the other hand, at low temperatures,
convection becomes an important energy transfer mechanism, influencing the
photospheric temperature stratification and, thus, hydrogen line
formation. The derived value of \Teff\ then depends on the convection
treatment used in the model atmosphere calculations.

There are basically two formalisms of convection treatment available for
model atmosphere computations: the well known mixing-length theory (MLT)
\citep{mlt}, which relies on some free parameter ($\alpha$), describing the
characteristic length traveled by convective cells before they disappear,
and further improvement of stellar convection developed by
\citet[][CM hereafter]{cm1}. The main advantage of the new model is
that it does not require any adjustable parameters like $\alpha$,
which is now computed based on the geometrical depth scale inside the
convective zone. It also accounts for the full spectrum of turbulent
eddies, and is therefore superior to the single-eddy assumption made by MLT.
More details can be found in \citet{cm2}.

Physically the CM convection model is somewhat superior to MLT, therefore
we preferred CM convection over MLT in our analysis. In the case of WASP-12,
we find no critical differences in H$\alpha$ profiles using both CM and MLT
convection theories with the commonly accepted $\alpha=1.25$. This is
because the wings of H$\alpha$ are formed mainly in the region right above
the photosphere where convection plays a less important role in the energy
balance. However, bluer Balmer lines are formed deeper and thus can show
the changes in the temperature stratification introduced by convection.
For instance, this is the case for the H$\gamma$ line for which MLT
predicts weaker line wings (i.e. stronger convective energy transport): one
needs an approximately $100$\,K hotter model to fit the H$\gamma$ line
wings with MLT convection. However, it is then clear that only CM allows
a consistent fit simultaneously for H$\alpha$ and H$\gamma$ lines with the
same \Teff.
\subsection{The distance and age of WASP-12}\label{age}
The basic parameters of WASP-12 were derived by \citet{hebb}. \Teff, \logg,
metallicity, and \vsini\ were derived from the observed stellar spectrum
using spectral synthesis; a simultaneous Markov chain Monte Carlo (MCMC)
fit to radial velocity and transit light curve measurements produced values
for $R_{\rm P}/R_*$, $R_*$, $M_*$ and the orbital semi-major axis $a$.
The observed spectral type of the star and its radius imply the luminosity
and the spectral energy distribution. Using this information, the 2MASS IR
fluxes can be used to derive the distance \citep[IRFM;][]{blackwell}. This
method yields a distance to WASP-12 of 265$\pm$20\,pc
({\tt http://www.superwasp.org/wasp\_planets.htm}), assuming a typical
main-sequence stellar radius; and 385$\pm$30\,pc, assuming the stellar
radius from the MCMC fits of \citet{hebb} (B. Smalley, priv.comm.). We note
that the IRFM implicitly assumes solar metallicity and zero reddening, both
inapplicable for WASP-12. Since the orbital period and the radial velocity
amplitude fix the value of $a$, and the transit light curve fixes the ratio
$a$:$R_*$, the stellar distance of 385$\pm$30\,pc is to be preferred.
\citet{hebb} also estimated the age of WASP-12 by fitting isochrones to the
position of WASP-12 in a modified HR diagram (temperature vs. inverse cube
root of the stellar density), obtaining an age of 2.0$^{+0.5}_{-0.8}$\,Gyr old;
given the several uncertainties, they increased the error bars to 1\,Gyr,
concluding that WASP-12 is 2$\pm$1\,Gyr old.

\begin{figure}
\begin{center}
\includegraphics[width=\hsize,clip]{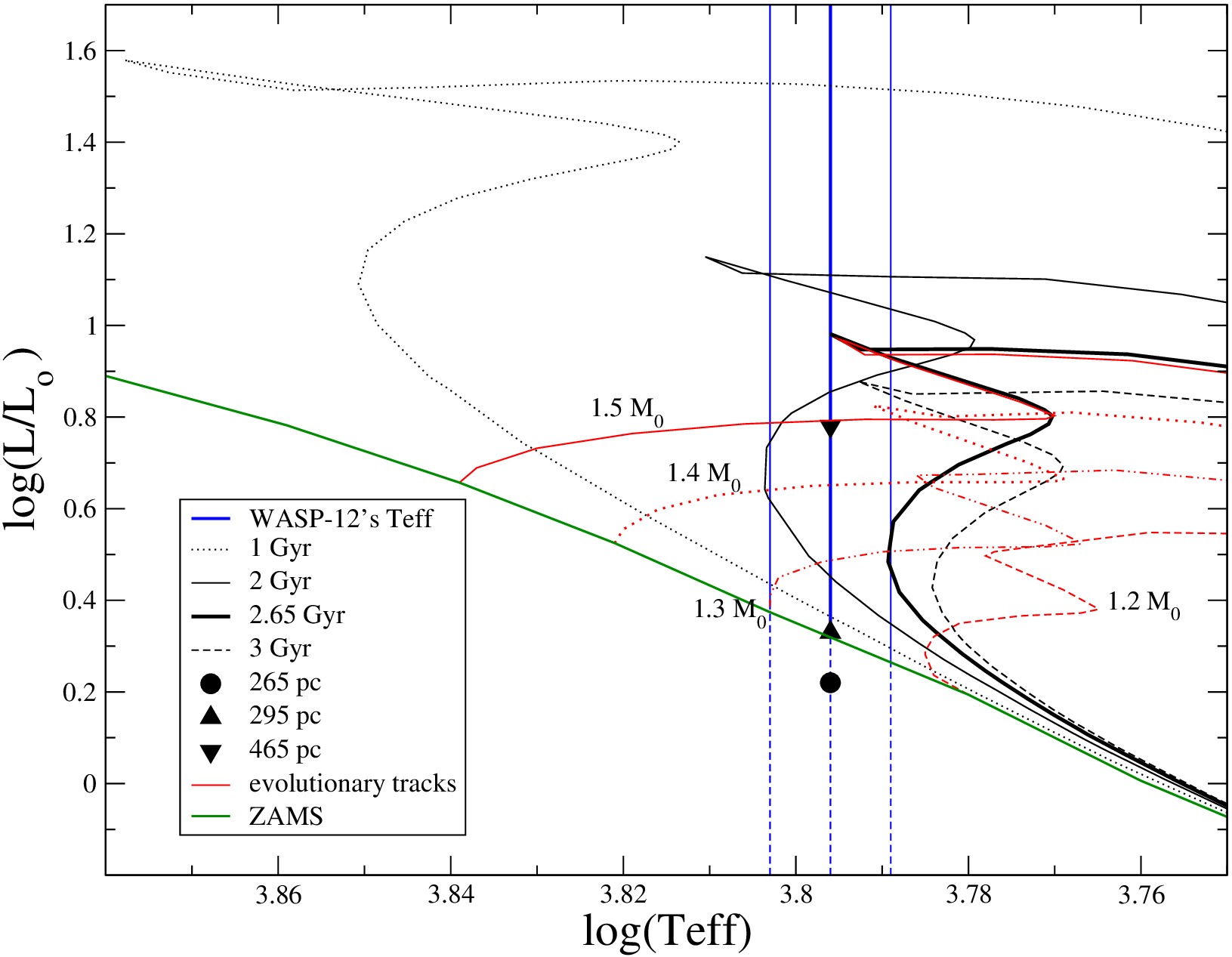}
\caption{Position of WASP-12 on the HR diagram assuming three different
stellar distances: 265\,pc (circle), 295\,pc (triangle),  and 465\,pc 
(inverted triangle). The maximum and minimum distances were calculated
adopting \Teff=6250\,K and interstellar reddenings from \citep{AL05}. The 
dotted, thin full and dashed lines show isochrones from \citet{marigo} 
corresponding to ages of 1\,Gyr, 2\,Gyr and 3\,Gyr, respectively, 
encompassing the possible age range of WASP-12 from \citet{hebb}. 
The thick full line is the 2.65\,Gyr isochrone we argue this is the 
maximum possible age for WASP-12. The red lines show evolutionary tracks
from \citet{girardi00} for  1.5\,\M, 1.4\,\M, 1.3\,\M and 1.2\,\M, from 
top to bottom. Both isochrones and evolutionary tracks assume a 
metallicity $Z$ of 0.03. The blue vertical lines show the WASP-12's 
temperature range; these lines change from full to dashed below the ZAMS, 
indicated by the green line.}
\label{fig.hr}
\end{center}
\end{figure}

One of the most secure empirical properties of the star is its colour, or
equivalently effective temperature. We can assess the distance and age of
WASP-12 by comparing the effective temperature with isochrones and
evolutionary tracks. Figure~\ref{fig.hr} shows four isochrones from
\citet{marigo} with a metallicity $Z$ of 0.030, the maximum available value
of $Z$, adopted following Eq.~\ref{isoZ}. WASP-12's effective temperature
places it on the vertical thick blue line of Fig.~\ref{fig.hr}, where the
other two full vertical blue lines are defined by the uncertainty on \Teff.
On this central line we placed three points, which correspond to distances
of 265\,pc (lower point), 295\,pc (middle point) and 465\,pc (upper point).
The lower point lies well below the zero-age main sequence (ZAMS). From
this we can rule out a distance as close as 265\,pc; this concurs with
the MCMC fitting of \citet{hebb} in implying the star is bigger than the
typical main sequence stellar radius.

We can see that a range of possible ages and distances are compatible with
the isochrones. In the region of Fig.~\ref{fig.hr} between the two solid
triangles, several isochrones are consistent with the empirical effective
temperature. The 2.65\,Gyr isochrone is just consistent with the lower limit
on WASP-12's effective temperature: the loop to the right at the main
sequence turn-off point of this isochrone intersects the lower limit on the
temperature. No star of this age is consistent with WASP-12's colour except
for those which are turning on to the horizontal branch, higher up in the
diagram at \logl$\approx$1. For stars at this stage of evolution
\logg$\approx$3.8, whereas the spectrum of WASP-12 implies \logg$\approx$4.2
(c.f. Sect~\ref{parameters}). WASP-12 thus cannot be turning on to the
horizontal branch. Consequently we can constrain the position of WASP-12
in Fig.~\ref{fig.hr} to be around or younger than the main sequence turn-off.

Applying this reasoning, the oldest possible age for WASP-12 is 2.65\,Gyr.
This arises from the intersection of lower limit on WASP-12's effective
temperature and the full thick isochrone in Fig.~\ref{fig.hr}. For all
isochrones younger than this, there is also an intersection at or before
the main-sequence turn-off. For the isochrones older than this no allowed
intersection occurs: only evolved stars have compatible temperatures, but
these are ruled out by their surface gravity.

The full red lines in Fig.~\ref{fig.hr} are evolutionary tracks, from
\citet{girardi00}, for stars of mass 1.2\,\M, 1.3\,\M, 1.4\,\M\ and 1.5\,\M\ 
respectively from bottom to top. WASP-12 is clearly hotter than a 1.2\,\M\ 
star for any age, therefore we conclude its mass exceeds 1.2\,\M.
Interpolating between the evolutionary tracks, we estimate a limit on the
mass of WASP-12 of around 1.23\,\M. This is consistent with MCMC fitting of
\citet{hebb}, but is a tighter constraint.

The evolutionary track for 1.3\,\M\ almost exactly coincides with the
intersection of the 2.65\,Gyr isochrone with the lower limit on the effective
temperature. The oldest possible age therefore corresponds to a stellar mass
of about 1.3\,\M. For higher masses, the empirical effective temperature
intersects the evolutionary track while the star is still on the main
sequence, and is therefore consistent with WASP-12's \logg. The 1.4\,\M\ 
evolutionary track is consistent with the empirical effective temperature,
and is also consistent with the MCMC fitting of \citet{hebb}.

The upper limit on the mass from \citet{hebb} is just below 1.5\,\M. We can
use this to infer an upper limit on the luminosity, and hence the distance.
We find that the maximum distance is 465\,pc. On the other hand, the minimum
distance is obtained at the intersection of the vertical line that defines
WASP-12's effective temperature and the ZAMS; this distance is 295\,pc. These
distances are computed taking into account insterstellar reddenings as given
by \citep{AL05}.

One effective way to measure the age of a late-type star is the comparison of
the lithium abundance with that of open cluster member stars for which the
age is precisely known.
For a more thorough analysis of the \ion{Li}{1} line at $\lambda$\,6707\,\AA\ 
we downloaded from the SOPHIE archive\footnote{http://atlas.obs-hp.fr/sophie/}
all the 21 mid-resolution spectra\footnote{SOPHIE is a cross-dispersed
\'{e}chelle spectrograph mounted at the 1.93-m telescope at the Observatoire
de Haute-Provence (OHP).} (R$\sim$40\,000) of WASP-12 obtained for the
radial velocity analysis published by \citet{hebb}.

\begin{figure}
\begin{center}
\includegraphics[width=13cm,clip]{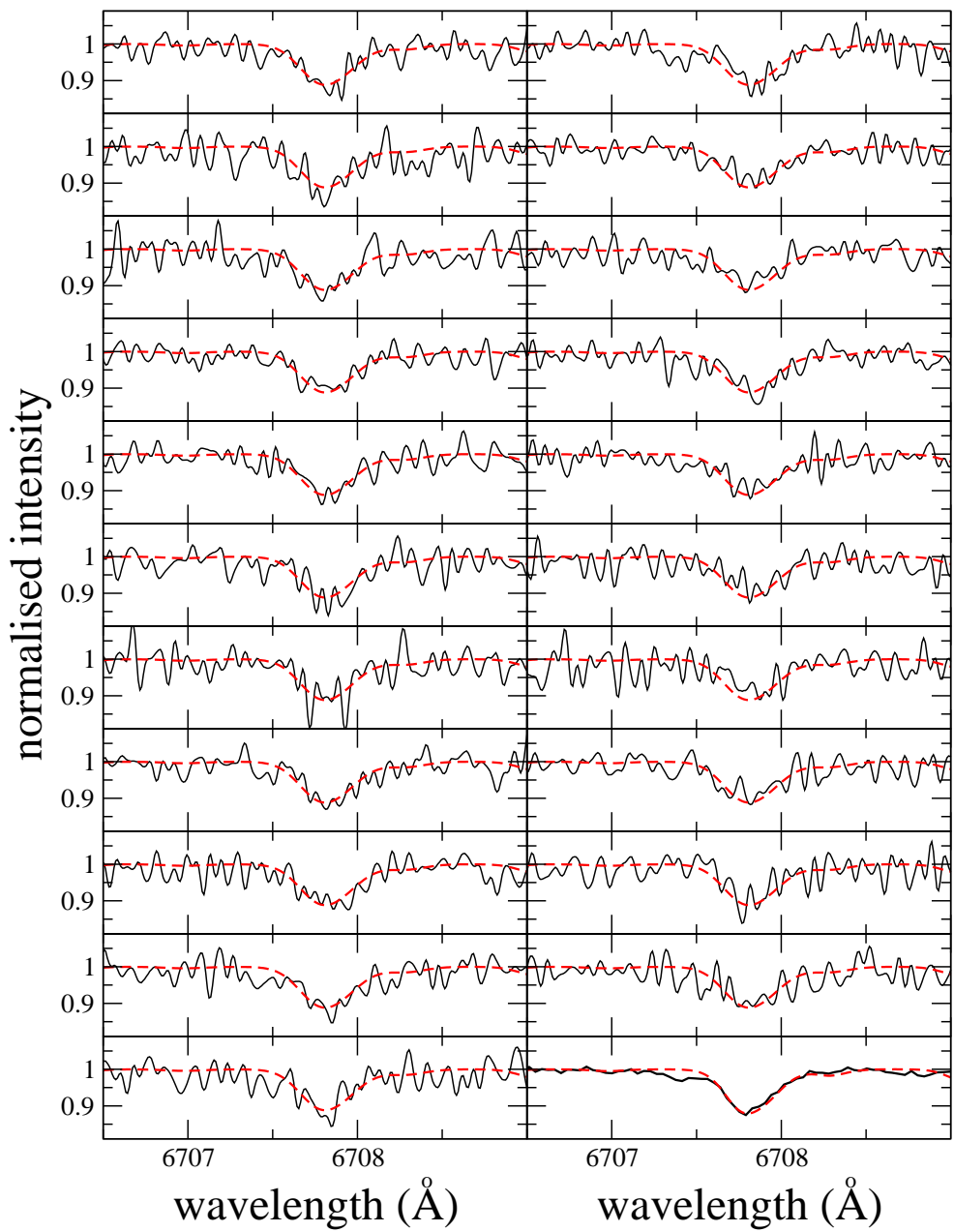}
\caption{Comparison between the observed phase dependent SOPHIE spectra
(black solid line) of WASP-12 around the \ion{Li}{1} line at
$\lambda$\,6707\,\AA\ and synthetic spectra (dashed red line) calculated with
the Li abundance and the parameters obtained from the analysis of the \espa\ 
data. The bottom-right panel shows the same comparison, but with the analysed
\espa\ spectrum instead. The synthetic spectra take into account the
difference in resolution between SOPHIE and \espa.}
\label{fig.lithium}
\end{center}
\end{figure}

Figure~\ref{fig.lithium} shows a comparison between the 21 SOPHIE spectra and
synthetic spectra calculated with the adopted stellar parameters and
abundances around the region of the \ion{Li}{1} line at $\lambda$\,6707\,\AA;
the bottom-right panel shows the same comparison, but with the \espa\ spectrum.
This plot shows that the Li abundance derived from the \espa\ spectrum fits
the SOPHIE data as well and that there is no line profile variation of the Li
line with the orbital phase. We measured also the equivalent width of the Li
line in each of these spectra and we did not find any significant time
variation. For this reason we believe that the lack of detection of this
\ion{Li}{1} line in the SARG spectrum of WASP-12, reported by \citet{hebb},
could be due to the low SNR or to a wrong line identification.

From a comparison between the lithium abundance of WASP-12 and the results
published by \citet{sestito} we can only conclude that WASP-12 is older
than 500\,Myr, which is in agreement with the 2.0$\pm$1.0\,Gyr given
by evolutionary tracks \citep{hebb}. The large uncertainty is caused by the
extremely slow lithium depletion for stars with effective temperatures
similar to that of WASP-12, due to the shallow surface convection zone.

In conclusion our analysis on the age of WASP-12 leads to a stellar age between
1.0\,Gyr \citep[lower limit given by][]{hebb} and 2.65\,Gyr (from the analysis
of the HR diagram), as shown also in Table~\ref{stellar.param} that lists the
minimum and maximum values we obtained for the stellar age, mass and distance.
\begin{table}[h]
\caption[ ]{Minimum and maximum values of WASP-12's age (in Gyr), \M\ 
and distance (in pc) as derived by the analysis of the HR diagram. The values 
marked with an ''*" are taken from \citet{hebb}.}
\label{stellar.param}
\begin{center}
\begin{tabular}{l|c|c}
\hline
\hline
Parameter & Minimum & Maximum \\
	  & value   & value   \\
\hline
age (Gyr)     & 1.0* & 2.65 \\
\M            & 1.23 & 1.49*\\
distance (pc) & 295  & 465  \\                                                                                                   
\hline
\end{tabular}
\end{center}
\end{table}
%
\subsection{Is WASP-12 a chemically peculiar star?}\label{chem.pec}
One of the main purpose of this work is to search for chemical
peculiarities that could be connected with pollution of the stellar
atmosphere by material lost by the planet. This can be done in
different ways, as also shown in the extensive salient literature present
\citep[see e.g.][and references therein]{neves2009}. WASP-12 is a promising
target for signs of atmospheric pollution: WASP-12\,b is most likely
currently losing material \citep{fossati2010}; this material is believed to be
forming a circumstellar disk that is accreting onto the star, polluting the
stellar photosphere \citep{li2010}. Classical ways of looking for atmospheric
pollution are by searching: ({\it i}) chemical peculiarities of single
elements, such as Li and Be \citep[e.g.][]{israelian2004}; ({\it ii}) a trend
in the element abundance against the condensation temperature
\citep[T$_c$; see e.g.][]{sadakane02,ecuvillon06,melendez09}; ({\it iii})
chemical peculiarities of the abundance pattern in comparison to reference
stars (both planet hosting and non-planet hosting).

When searching for small effects on the stellar atmospheric abundances
it is important to determine whether a certain star belongs to the thin or
thick Galactic disk population. To do so we performed both a kinematic and
chemical analysis of WASP-12. We calculated the Galactic
velocity vectors (U, V, W) corrected to the local standard of rest (LSR)
using the formalism of \citet{johnson87}, instead defining U as positive
towards the Galactic anti-centre. As done by \citet{sozzetti2006}, WASP-12 was
then placed on the Toomre diagram of the \citet{soubiran} stellar sample,
indicating that WASP-12 has a peculiar velocity less than 85\,\kms, strongly
indicative of thin disk membership. From the WASP-12 abundances, we then
compared our values of [$\alpha$/Fe] ($-$0.15\,dex) and [Fe/H] ($+$0.28\,dex),
where [$\alpha$/Fe] is defined as 0.25([Mg/Fe]+[Si/Fe]+[Ca/Fe]+[Ti/Fe]), with
the \citet{soubiran} sample, obtaining that WASP-12 is again consistent
with the properties of the thin disk population.

In our analysis we are not able to look for chemical peculiarities of both Li
and Be. As shown in Fig.~\ref{fig.lithium}, our spectra do not have the SNR
and the resolution necessary to perform a precise analysis of
the Li$^6$/Li$^7$ ratio, but we can compare the lithium abundance obtained for
WASP-12 with that of other planet hosting stars. \citet{israelian2004} and
\citet{gonzalez2010a} published lithium abundances of stars hosting a giant
planet or a brown dwarf and compared them to a set of non-planet hosting stars.
The lithium abundance we obtained for WASP-12 matches that of both set of
stars, showing clearly that Li is not peculiar in WASP-12. Our spectra do
not include the region around $\lambda$\,3130\,\AA\ covering the two
\ion{Be}{2} lines usually adopted to measure the Be abundance.
\subsubsection{Volatile vs. refractory elements}
\begin{figure}
\begin{center}
\includegraphics[width=\hsize,clip]{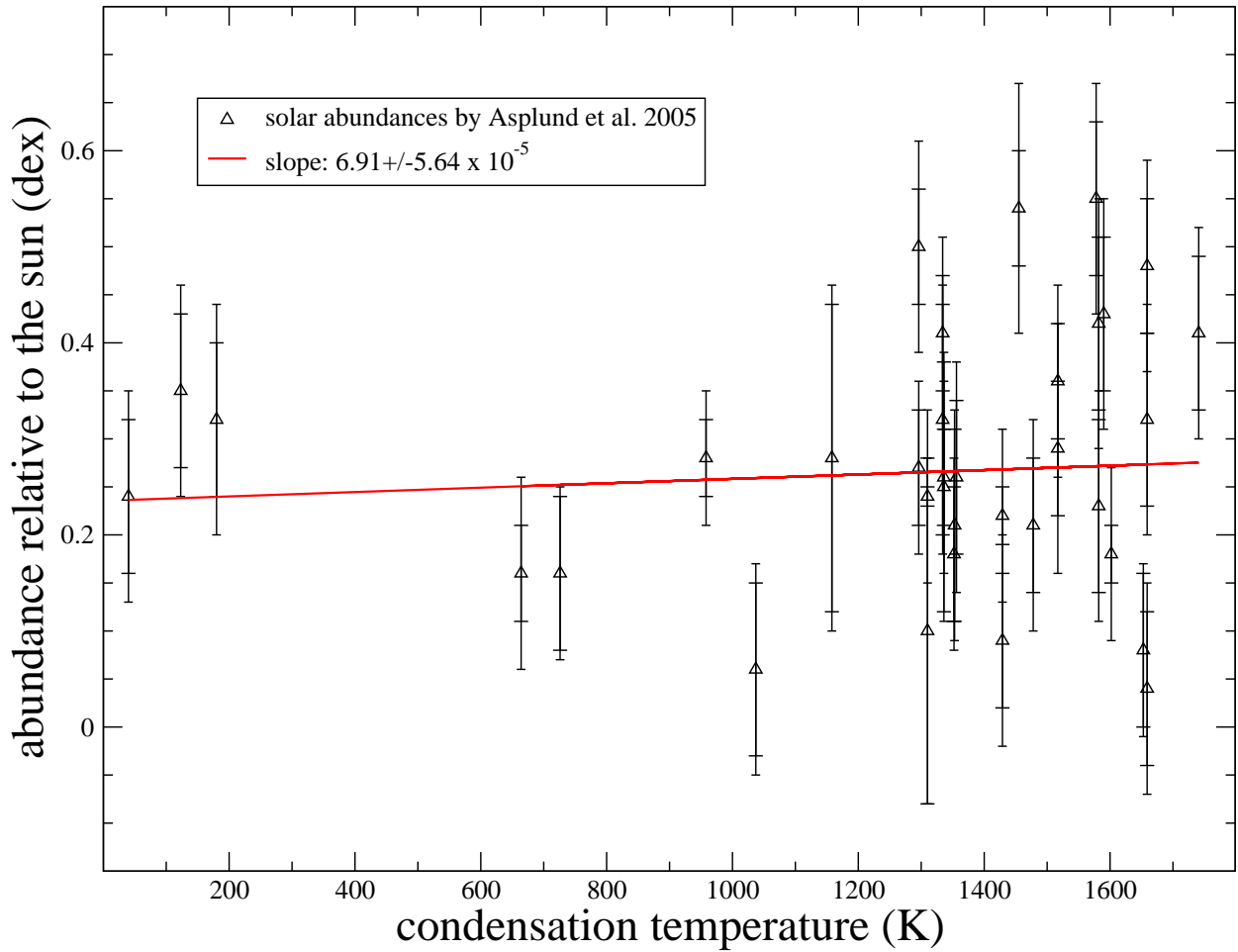}
\caption{Ion abundance relative to the sun \citep{met05} as a function of
the condensation temperature. The full line shows the linear fit to the data.
The abundance uncertainties are given as in Fig.~\ref{fig.abn}.}
\label{fig.condtemp}
\end{center}
\end{figure}

An effective way to check whether a stellar photosphere is polluted by
accretion of metal-rich material is to examine the correlation between the
relative abundance of various elements and their T$_c$. In accreting
metal-rich material the refractory elements tend to form dust grains that
are blown away by the stellar wind. Thus the star accretes more
volatile than refractory elements, as happens, for example, in
$\lambda$~Bootis stars \citep{bootis}. Our WASP-12 ion abundance are shown
against condensation temperature \citep[taken from][]{lodders} in
Fig.~\ref{fig.condtemp}.

The correlation obtained between the ion abundances relative to the Sun
\citep{met05} and T$_c$ is not statistically significant:
6.91$\pm$5.64$\times$10$^{-5}$. In the past years several authors
\citep[e.g.][]{sadakane02,ecuvillon06,sozzetti2006} looked for the
same kind of correlation in several planet hosting stars, without having found
any. The trend we obtain for WASP-12 is also indicative of a null result.

\citet{melendez09}, applying a differential analysis on the Sun and solar
twins, discovered that the Sun shows a highly significant correlation of the
abundances (relative to the mean abundance of the solar twins) in respect to
the condensation temperature. In addition this correlation shows a significant
break at T$_c\sim$1200\,K. \citet{melendez09} and \citet{gonzalez2010b} came
to the conclusion that this correlation is likely connected to the presence
of planets around the Sun. The detection of this kind of correlation requires
very high precision abundances (error bars $<$~0.05\,dex and free from
systematic differences), attainable only with a differential analysis. The
application of this method to WASP-12 would require observations of stars
that can be considered twins of WASP-12 (similar \Teff, metallicity, age,
and population). The choice of the WASP-12 twins will be of crucial
importance because effects such as diffusion, more prevalent for F-type
stars than G-type stars, are strongly age and \Teff\ dependent, and
consequently could easily hide or mimic pollution signatures.

Our analysis does not lead to a firm conclusion about the atmospheric
pollution of WASP-12.
\subsubsection{Comparison with previous analysis}
To check whether WASP-12 has a peculiar abundance pattern we compared the
abundances with those obtained by other authors on a large number of stars
with a similar \Teff. In particular we took into account the
results of the abundance analysis on stars that are not known planet hosts,
in the temperature range 6000--6500\,K, and with a [Fe/H]$>$0.0\,dex. We
decided to use such a small temperature range to decrease the effect of
possible systematics, such as \nlte\ effects. In addition, since
we are interested in looking for a peculiar pattern, we decided to use only
the stars with an over-solar iron abundance and to subdivide the sample of
comparison stars according to their iron abundance: 0.0$<$[Fe/H]$\leq$0.1,
0.1$<$[Fe/H]$\leq$0.2, and [Fe/H]$>$0.2. To have a better statistical view of
each subsample we decided to plot in our comparisons not only the mean
abundance and relative standard deviation for the comparison stars, but also
their abundance range. To accomplish all these requirements, we
needed to have a large set of non-planet hosting stars, such as the one
provided by \citet{valenti}.

\begin{figure}
\begin{center}
\includegraphics[width=\hsize,clip]{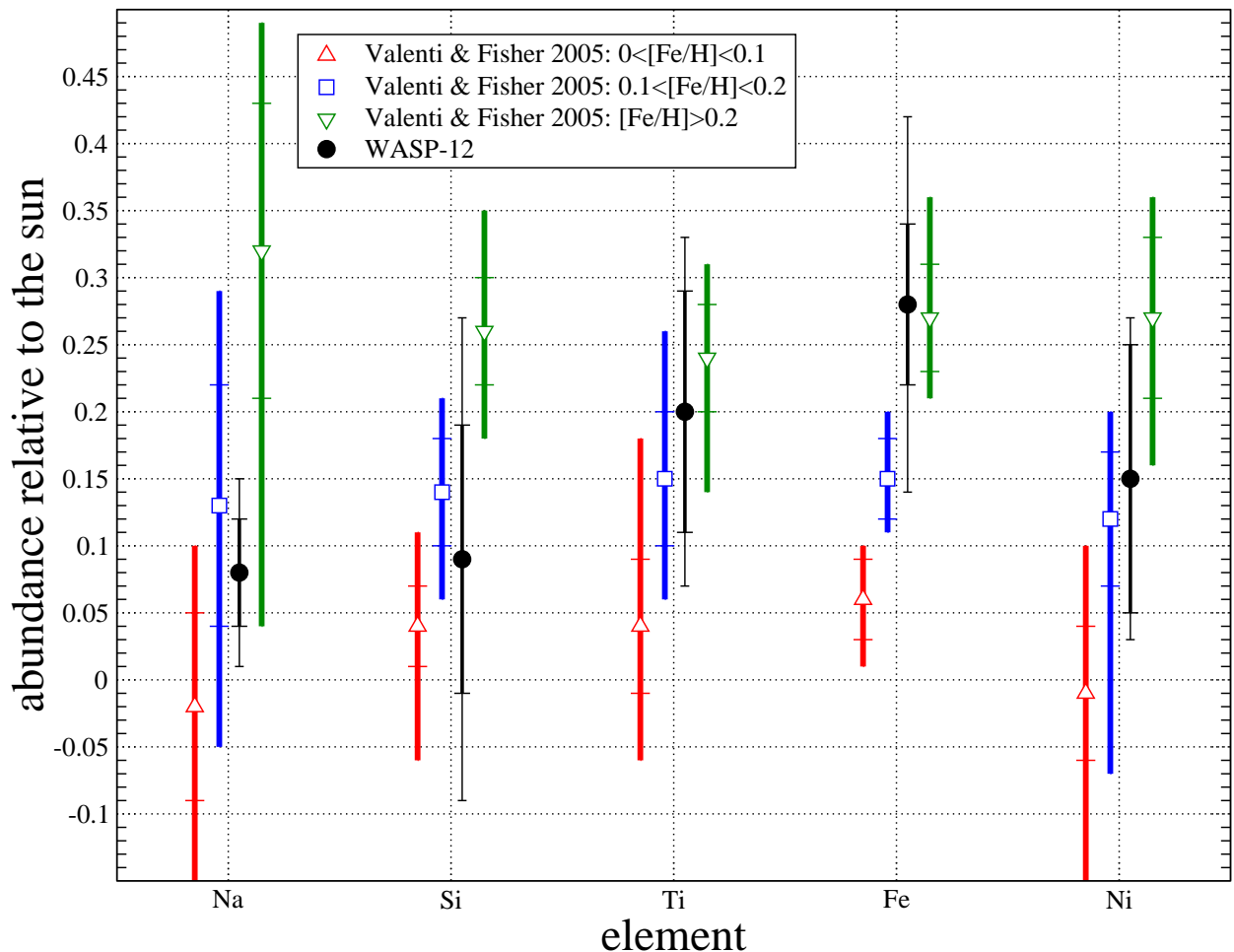}
\caption{Comparison between the mean abundance, relative to the sun, of
WASP-12 (full circles) and of the selected sample of stars from
\citet{valenti}, that was subdivided according to the iron abundance:
0.0$<$[Fe/H]$\leq$0.1\,dex (open upside triangles), 0.1$<$[Fe/H]$\leq$0.2\,dex
(open squares), and [Fe/H]$>$0.2\,dex (open downside triangles). The shaded
areas show the abundance range, while the error bars of the comparison
sample give the standard deviation from the mean abundance. The uncertainties
on the abundances of WASP-12 are as in Fig.~\ref{fig.abn}.}
\label{fig.valenti}
\end{center}
\end{figure}

\citet{valenti} performed a parameter determination and LTE abundance
analysis of more than 1000 late-type main sequence stars. For each star they
derived the abundances of Na, Si, Ti, Fe, and Ni. This sample provided us with
comparison abundances for 59 stars with 0.0$<$[Fe/H]$\leq$0.1\,dex, 36 stars
with 0.1$<$[Fe/H]$\leq$0.2\,dex, and 22 stars with [Fe/H]$>$0.2\,dex.
The comparison between our WASP-12 abundances and the
ones from the sample of \citet{valenti} is shown in Fig.~\ref{fig.valenti}.

Figure~\ref{fig.valenti} shows that the abundances of Ti and Fe are clearly
comparable to the ones of the higher metallicity subsample, as expected given
the fact that [Fe/H] of WASP-12 is about 0.3. The Na abundance of
WASP-12 falls within the abundance range of the high-metallicity stars of the
comparison sample, due to the large range covered by the Na abundance,
while Ni shows a mean abundance that is just outside the abundance range.
The Si abundance looks definitely more similar to the one of the set of stars
with 0.1$<$[Fe/H]$\leq$0.2\,dex, but the large uncertainty on this abundance
does not allow any firm conclusion. In summary the abundances of Ti and Fe
follow the pattern of the high-metallicity stars, while Na, Si, and Ni seem
to follow more the abundance pattern of the set of comparison stars with
[Fe/H] between 0.1 and 0.2\,dex.

The information gathered from this comparison lead to the conclusion
that there is the possibility that the WASP-12 abundances of Na, Si, and Ni
do not really match the abundance pattern that the WASP-12 iron abundance
would suggest. But it has to be taken into account that the comparison is
not free from systematic differences in the fundamental parameters and
abundance determinations, and in the set of adopted atomic line parameters.

For a further search of possible abundance peculiarities in the atmosphere of
WASP-12 we decided to compare our results with those of several other authors
\citep{takeda02,sadakane02,ecuvillon04,sozzetti2006,gilli06,santos06,bond08,neves2009}
have obtained in planet hosting and non-planet hosting stars. In this way
we should also be able to partially remove possible systematic differences in
case a peculiar pattern becomes evident in the comparisons with several
different authors.

The WASP-12 abundances we derived for O, Ca, Sc, Mn, Zn, Y, Zr, Nd, and Eu
match very well the ones previously obtained by other authors in planet hosting
and non-planet hosting stars, while abundances of Al, V, and Cu appear
to be clearly below the values previously obtained by more than 0.2\,dex.
For the other compared elements (C, N, Na, Mg, Si, S, Ti, Cr, Co, and Ni)
we obtained just a satisfactory agreement, since we register a tendency of the
WASP-12 abundances of these elements to lay always in the lower margin of the
comparison samples, in particular when the abundances are put in relation
to the iron abundance. This result follows what previously obtained,
strengthening the possibility of an increased Fe abundance in comparison
to the one of some other elements.

In addition we compared our Si, Ti, and Ni WASP-12 abundances with the ones
obtained by \citet{robinson2006} in a set of planet hosting and non-planet
hosting stars. We obtained a good agreement for Ti, but both Si and Ni appear
to be depleted in WASP-12, compared to the abundances obtained by
\citet{robinson2006} in a set of planet hosting stars, therefore we cannot
confirm their conclusion of a systematic enrichment of Si and Ni in planet
hosting compared to non-planet hosting stars.
\subsection{Spectral energy distribution: searching for a circumstellar disk}
\label{spec.energy}
\begin{figure}
\begin{center}
\includegraphics[width=\hsize,clip]{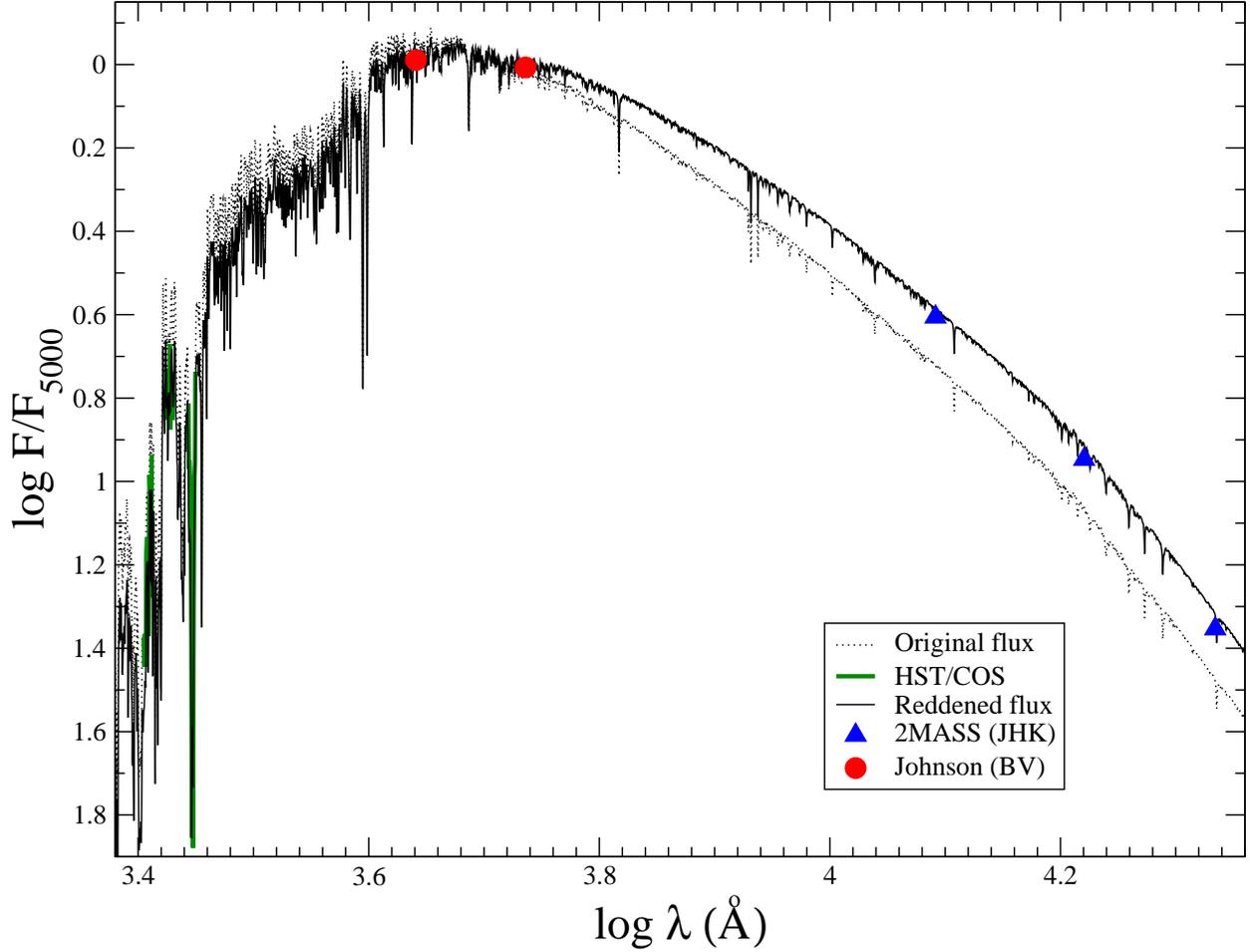}
\caption{Comparison between \llm\ theoretical fluxes calculated with the
fundamental parameters and abundances derived for WASP-12, taking into
account a reddening of E(B-V)=0.126\,mag (full line) and without
taking into account reddening (dashed line), with COS calibrated fluxes
(thick gray line), Johnson $BV$ photometry (full points) and 2MASS photometry
(full triangles). The error bars on the photometry have the same size of the
symbols. The model fluxes and the COS spectra were convolved to have
approximately a spectral resolution of 800, for display reasons.}
\label{fig.flux}
\end{center}
\end{figure}

The circumstellar disk, predicted by \citet{li2010} and tentatively observed by
\citet{fossati2010}, could be detectable in the infrared. For this reason we
decided to compare the calibrated Near-UV COS fluxes of WASP-12 and the
available photometry (Johnson and 2MASS photometry) with synthetic fluxes
obtained with \llm\ adopting the fundamental parameters and the abundances
derived from the \espa\ data, looking for an infrared excess in the 2MASS
photometry. For this comparison, shown in Fig.~\ref{fig.flux}, we took into
account a reddening E(B-V)=0.126\,mag \citep{AL05}, that was calculated
assuming the stellar distance obtained in Sect.~\ref{age}.

Figure~\ref{fig.flux} shows a very good agreement between the model fluxes and
the observations, demonstrating also the quality of both the adopted
fundamental parameters and interstellar reddening.

Figure~\ref{fig.flux} does not show any clear infrared excess in the 2MASS
photometry that could be interpreted as: ({\it i}) there is no interstellar
disk around the star or ({\it ii}) the interstellar disk is not visible
because either the emission is not strong enough or the emission peaks at
much longer wavelengths. \citet{li2010} suggested that the disk emission
should peak at $\sim$2.3\,$\mu$m, inside the wavelength range covered by the
$K$ band 2MASS photometry. Therefore, to look for an interstellar disk around
WASP-12 it would be necessary to observe the system with a very high precision
\citep{li2010} and at longer wavelengths, both accessible with the Herschel
satellite \citep{herschel}.
\section{Conclusions}\label{conclusions}
On January 3rd and 5th 2010 we observed the planet hosting star WASP-12 with
the \espa\ spectropolarimeter with the aim of looking for a global magnetic
field, derive the stellar parameters and perform a precise abundance analysis.

The WASP-12 fundamental parameters and iron abundance we obtained are in
agreement with what previously published by \citet{hebb}:
\Teff=6250$\pm$100\,K, \logg=4.2$\pm$0.2, \vmic=1.2$\pm$0.3\,\kms, and
[Fe/H]=0.32$\pm$0.12\,dex. \vsini\ is less than 4.6$\pm$0.5\,\kms,
while \vmac\ lays within the range 4.75--7.0$\pm$0.6\,\kms. Dedicated
observations to measure the amplitude of the RM effect are necessary to derive
the real stellar \vsini. The resultant metallicity of WASP-12 is
$Z$=0.021$\pm$0.002\,dex. A detailed analysis of the HR diagram, with the use of
isochrones and evolutionary tracks allowed to derive more accurate ranges for
the stellar age, mass and distance: the age of WASP-12 is between 1.0\,Gyr
\citet{hebb} and 2.65\,Gyr, the mass is between 1.23 and 1.49\,\M\ 
\citep[the last value comes from][]{hebb}, and the distance to WASP-12 is
between 295 and 465\,pc. Our measurement of the Li abundance allowed us just
to conclude that WASP-12 is older than 500\,Myr.

We performed a magnetic field search adopting the LSD technique revealing
that the star does not show any magnetic field signature in Stokes $V$.
A detailed analysis of the possible star-planet magnetic interaction
would require a time-dependent analysis of particular spectral lines,
such as the Ca H \& K lines \citep{es03,es05,es08} and the \ion{Mg}{2} UV
resonance lines, but it has to be taken into account that WASP-12 shows a
remarkably low stellar activity \citep{knutson}, that will be analysed in
detail in a forthcoming paper.

Given recent theoretical predictions \citep{li2010} and discoveries
\citep{fossati2010}, the WASP-12 system seems to be an ideal target to detect
the presence of atmospheric pollution, due to the material lost by the planet.
Therefore we looked for hints of pollution by looking for a correlation between
the atmospheric element abundances and the condensation temperature, and by
comparing the WASP-12 abundance pattern with the one of other planet hosting
and non-planet hosting stars, previously published by several other authors.
Our analysis revealed just the presence of hints of atmospheric pollution,
although only a differential analysis would allow to obtain firm evidences.
One must also take into account the fact that it is not clear whether WASP-12
would show the same kind of atmospheric pollution shown by the sun
\citep{melendez09}: the material coming from WASP-12\,b and falling onto
the star is in a gas/plasma state and only a detailed modeling of the
temperature and density structure of the accretion disk would show whether
the material is condensing in dust grains. If dust grains are forming it is
likely that a differential analysis would reveal the kind of pollution
signatures obtained for the sun, otherwise all the material lost by the
planet would fall onto the star, making then the pollution signature
dependent to the unknown hydrogen content of the planet.

With the use of the available HST calibrated spectra and of visible and
infrared photometry, we looked for the presence of a circumstellar disk around
WASP-12, but without success. Probably high precision far infrared
measurements, such as those possible with the Hershel satellite, may
reveal the presence of a circumstellar disk.
\section*{Acknowledgments}
This work is based on observations obtained at the Canada-France-Hawaii
Telescope (CFHT), which is operated by the National Research Council of
Canada, the Institut National des Sciences de l'Univers of the Centre
National de la Rechereche Scientifique of France, and the University of
Hawaii. This work is also based on observations made with the NASA/ESA Hubble
Space Telescope, obtained from MAST at the Space Telescope Science Institute,
which is operated by the Association of Universities for Research in
Astronomy, Inc., under NASA contract NAS 5-26555. These observations are
associated with program \#11651 to which support was provided by NASA through
a grant from the Space Telescope Science Institute. This publication makes use
of data products from the Two Micron All Sky Survey, which is a joint project
of the University of Massachusetts and the Infrared Processing and Analysis
Center/California Institute of Technology, funded by the National Aeronautics
and Space Administration and the National Science Foundation. This work is
supported by an STFC Rolling Grant (L.F., A.E.). O.K. is a Royal Swedish
Academy of Sciences Research Fellow supported by grants from the Knut and
Alice Wallenberg Foundation and the Swedish Research Council. D.S. acknowledges
support received from the Deutsche Forschungsgemeinschaft (DFG) Research
Grant RE1664/7-1. L.F. is deeply in debt with Dr. Tanya Ryabchikova for
the tremendous help she gave with the \loggf\ values' references,
Dr. Gregg Wade for his help during the preparation of the observing proposal,
and Dr. Berry Smalley for the information regarding the determination of the
WASP-12 distance. We thank also the CFHT staff, Dr. Nadine Manset and the
CFHT director, Dr. Christian Veillet, for their help and speed in obtaining
the observations.
{\it Facilities:} \facility{HST (COS)}, \facility{CFHT (ESPaDOnS)},
\facility{OHP (SOPHIE)}.

\end{document}